\documentstyle[12pt, epsf]{article}
\newcommand\ZZZ{{\hbox{ Z\kern-1.6mm Z}}}
\newcommand{\beq}{\begin{equation}}
\newcommand{\eeq}{\end{equation}}
\newcommand{\bea}{\begin{eqnarray}}
\newcommand{\eea}{\end{eqnarray}}
\newcommand{\ra}{\rangle}
\newcommand{\la}{\langle}
\newcommand{\lt}{\left}
\newcommand{\rt}{\right}
\newcommand{\Iop}{\relax{\rm I\kern-.18em I}}
\newcommand{\one}{{\hbox{ 1\kern-1.2mm l}}}

\newcommand{\del}{\partial}

\newcommand{\sectiono}[1]{\section{#1}\setcounter{equation}{0}}
\newcommand{\subsectiono}[1]{\subsection{#1}}

\begin{document}
{}~
{}~
\hfill\vbox{\hbox{DAMTP-2007-30}}
\break

\vskip 2cm

\centerline{\Large \bf A Universality in PP-Waves}

\medskip

\vspace*{4.0ex}

\centerline{\large \rm Partha Mukhopadhyay }

\vspace*{4.0ex}

\centerline{\large \it Department of Applied Mathematics and Theoretical Physics}
\centerline{\large \it University of Cambridge}
\centerline{\large \it Wilberforce Road, Cambridge CB3 0WA, UK}

\medskip

\centerline{E-mail: P.Mukhopadhyay@damtp.cam.ac.uk}

\vspace*{5.0ex}

\centerline{\bf Abstract} \bigskip

We discuss a universality property of any covariant field theory in space-time expanded around pp-wave backgrounds. According to this property the space-time lagrangian density evaluated on a restricted set of field configurations, called universal sector, turns out to be same around all the pp-waves, even off-shell, with same transverse space and same profiles for the background scalars. In this paper we restrict our discussion to tensorial fields only. In the
context of bosonic string theory we consider on-shell pp-waves and argue that universality requires the existence of a universal sector of world-sheet operators whose correlation functions are insensitive to the pp-wave nature of the metric and the background gauge flux. Such results can also be reproduced using the world-sheet conformal field theory. We also study such pp-waves in non-polynomial closed string field theory (CSFT). In particular, we argue that for an off-shell pp-wave ansatz with flat transverse space and dilaton independent of transverse coordinates the field redefinition relating the low energy effective field theory and CSFT with all the massive modes integrated out is at most quadratic in fields. Because of this simplification it is expected that the off-shell pp-waves can be identified on the two sides. Furthermore, given the massless pp-wave field configurations, an iterative method for computing the higher massive modes using the CSFT equations of motion has been discussed. All our bosonic string theory analyses can be generalised to the common Neveu-Schwarz sector of superstrings.

\newpage

\tableofcontents

\baselineskip=18pt

\sectiono{Introduction and Summary}
\label{s:intro}

In a previous work \cite{mukhopadhyay06} we argued that string theory in the maximally supersymmetric type IIB plane wave background with constant five-form Ramond-Ramond (RR) flux (RR plane wave) \cite{blau01} possesses a universality property. According to this property there exists a universal sector of string configurations for which the classical space-time
action evaluated around the RR plane wave takes the same value as in flat space. The argument leading to this property involved showing that the relevant light-cone world-sheet theory
\cite{metsaev01, metsaev02} has certain properties that are required to hold in order for the
universal sector to exist. By looking at the other examples of pp-waves where string theory has been studied in light-cone gauge, this universality property was then generalised to all the exact string solutions of pp-wave type, hereafter called {\it exact pp-waves}, with the same profile for the dilaton.

The original motivation for this work was to study open string tachyon condensation \cite{tachyon} in RR backgrounds. In fact, it was argued that the above universality property applied to the D-brane worldvolume actions has the following indication: For every D-brane configuration extending along the light-cone directions $x^{\pm}$ in flat space there exists a
corresponding D-brane configuration in each of the relevant pp-waves. Moreover for unstable configurations the tachyon potential is universal. The aforementioned correspondence is in the sense that the tachyon solutions representing such D-branes in the worldvolume theory on an unstable D-brane configuration are universal. This also implies the same D-brane descent relations of flat space \cite{sen99} in case of the relevant D-branes in pp-waves.

The above argument is based on the classical theory in space-time and therefore understanding of the systematics of the string loop expansion in RR backgrounds seems important in this case\footnote{I thank Ashoke Sen for emphasising this point.}. Assuming this reliability, if correct, the above results are non-trivial at least for the following reasons: The universality of tachyon potential is a new result for an RR background which does not follow from the analysis of \cite{universality}. Once we know how to formulate open string field theory in an RR pp-wave, we have a prediction for the tachyon potential. Moreover, the universal sector of D-branes contain non-BPS D-branes which are difficult to study using a manifestly supersymmetric formalism \cite{mukhopadhyay00, nemani, mukhopadhyay04,  mukhopadhyay05}. The open string boundary condition for the world-sheet fields that are space-time spinors were found in \cite{mukhopadhyay04} to be bi-local in flat background. This boundary condition should have proper generalisations to other relevant pp-waves\footnote{This question wa studied in \cite{mukhopadhyay06} for the case of type IIB RR plane wave.}.

Although the arguments of \cite{mukhopadhyay06}, which was designed to relate the space-time theory and the light-cone world-sheet theory, showed the existence of a universal sector of string configurations, its precise description was not found. It was also not clear whether
string theory plays any special role as long as the universality is concerned. The purpose of the present paper is to go beyond the ``light-cone argument'' of \cite{mukhopadhyay06} and study the universality property further with a fully covariant treatment. The results of this paper are summarised below:
\begin{enumerate}
\item
We consider an arbitrary covariant field theory of tensorial fields in space-time and define the universal sector of field configurations. We argue that the space-time lagrangian density evaluated around a pp-wave background, even off-shell, with the fluctuations restricted to the universal sector, is insensitive to the pp-wave nature of the background metric and background fluxes.
\item
Using the construction of bosonic non-polynomial closed string field theory (CSFT) \cite{csft, zwiebach92} in an exact pp-wave given in terms of the relevant world-sheet conformal field theory (CFT) we argue that universality requires the existence of a universal sector of world-sheet operators whose correlation functions are insensitive to the background gauge flux and the pp-wave nature of the background metric. Such requirements can indeed be verified
explicitly using the world-sheet CFT.
\item
Using the equivalence between the low energy effective field theory (EFT) and CSFT with all the higher massive modes integrated out, as formulated by Ghoshal and Sen in \cite{ghoshal-sen}, we describe a general method of identifying between the two sides the off-shell pp-wave configurations with flat transverse space and dilaton independent of the transverse coordinates. This method also includes an iterative procedure to compute all the massive field configurations using their equations of motion.
\end{enumerate}
Below we shall discuss these results in more detail following the same order.

A pp-wave \cite{brinkmann, duval91, tseytlin} admits a covariantly constant null Killing vector corresponding to translation of a particular coordinate $v$ (or $x^-$). Generically an $(N+2)$-dimensional pp-wave can be obtained by attaching an $N$-dimensional euclidean space (called the transverse space) parameterised by $x^I$ ($I=1,2,\cdots, N$) with a two-dimensional space parameterised by $u$ (or $x^+$) and $v$ such that the only non-zero component of the covariant metric tensor with a $v$ index is $\hat g_{uv}$, which is constant. In Brinkmann coordinates \cite{brinkmann} the pp-wave nature of the metric is manifested in the fact that the components $\hat g_{uu}(u,x)$ and $\hat g_{uI}(u,x)$ are generically non-zero, but independent of $v$ as required by the null Killing symmetry. The background can
also contain null gauge fluxes and scalars, all independent of $v$. The universal sector of field configurations is obtained by restricting the number of covariant $v$ indices with respect to the number of covariant $u$ indices of a generic tensorial fluctuation in a certain way and not letting the fluctuation vary along $v$.  By considering an arbitrary covariant field theory in space-time we argue that the effects of the background fluxes and the metric components $\hat g_{uu}$ and $\hat g_{uI}$ are erased inside the universal sector so that the effective metric becomes:
\bea
ds_{eff}^2= 2 du dv + G^{\perp}_{IJ}(u,x) dx^I dx^J~,
\label{Geff}
\eea
where $\hat g_{IJ}= G^{\perp}_{IJ}(u,x)$ is the transverse space metric which is also independent of $v$. This, however, is not true for the background scalar fields. Since the vacuum expectation value of a scalar field can be thought to determine the effective couplings of a theory, the universal sector should be sensitive to that value. This gives the complete covariant description of universality (for tensorial fields) as observed in \cite{mukhopadhyay06} in the context of light-cone gauge.

Given the above discussion, which directly concerns the space-time theory, we then proceed to study its relevance to string theory from the world-sheet point of view. Our discussion in the context of string theory will be restricted to bosonic strings which can be generalised to the common Neveu-Schwarz (NS) sectors of superstring theories. In this framework we deal with the world-sheet sigma model CFT corresponding to an exact pp-wave with dilaton independent of transverse coordinates \cite{amati89, horowitz90, rudd91, duval93}. A natural framework where
the above discussion of universality can be related to the world-sheet theory appears to be the closed string field theory (CSFT). Using the connection between CFT and CSFT\footnote{A CSFT is usually constructed with reference to a world-sheet CFT and it describes the space-time theory expanded around the corresponding background.}, we identify a set of operators in the CFT, called the universal sector of operators, which correspond to the universal sector of field configurations. We then argue that universality requires the correlation functions involving only such operators to be insensitive to the background gauge flux and the pp-wave nature of the background metric. More precisely, such correlation functions are expressed in terms of those in the effective background (\ref{Geff}) with both the gauge flux and dilaton switched off. We verify these relations explicitly using the world-sheet theory.

Finally we study pp-waves with flat transverse space and dilaton independent of transverse directions in CSFT. It was argued by Horowitz and Steif in \cite{horowitz90} that certain more restricted pp-waves which are solutions of the low energy effective field theory (EFT) are also exact solutions in string theory to all orders in $\alpha'$ expansion. By exploiting certain special geometrical properties of the backgrounds it was argued that all the $\alpha^{\prime}$ corrections to the EFT equations of motion drop off for these restricted pp-waves. This, however, is not true for the generic pp-wave ansatz of our consideration as described at the beginning of this paragraph. In this case one expects to receive corrections to all orders\footnote{It was argued in  \cite{rudd91} that such corrections can give rise to new
exact ``stringy" solutions.}. Certainly computing these corrections to arbitrarily high order seems impossible from the point of view of sigma model perturbation theory. Nevertheless, an encouraging feature is that all such corrections to the equation of motion are quadratic in
fields. Since in the context of CSFT $\alpha'$ does not play the role of a small expansion parameter, one may wonder if such corrections can be computed using this framework. This, however, requires us to find the precise relation between EFT and CSFT. This problem was formulated by Ghoshal and Sen in \cite{ghoshal-sen}, which we refer the reader to for the details. The basic idea behind this relationship is that after integrating out all the massive modes of CSFT by using their equations of motion one arrives at an effective field theory with the same massless dynamical degrees of freedom as in EFT such that it is related to EFT through suitable field redefinition. Although this field redefinition is quite complicated in general, once known, it provides us with a map between the off-shell configurations of EFT and CSFT with all the massive fields integrated out. We point out that a Horowitz-Steif type argument implies that for a generic off-shell pp-wave ansatz with flat transverse space and dilaton independent of transverse coordinates such field redefinitions simplify enormously to become at most quadratic in fields. Given this expectation we then proceed to formulate a CSFT computation that will enable us to find this field redefinition and therefore the off-shell map relating such pp-waves in EFT and CSFT. Moreover, it turns out that all the aforementioned $\alpha'$ corrections to the equation of motion are invariant under this field redefinition and therefore can be directly obtained from the relevant CSFT computation. This computation, however by itself, does not enable us to find the massive modes on the CSFT side given the profiles of the massless modes of EFT. To this end we discuss an iterative method of computing the massive modes of CSFT using their equations of motion. Therefore the whole construction enables us to
find the CSFT solution for any exact pp-wave, given its form on the EFT side, or equivalently, given the world-sheet sigma model CFT\footnote{See \cite{michishita} for an earlier discussion on this topic.}.

It would be interesting to perform the relevant CSFT computations to realise the above procedure explicitly and find the desired $\alpha^{\prime}$ corrections to the equation of motion. It would also be interesting to study the universality property by including the
space-time spinors. Although we do not have a formulation of string field theory in RR backgrounds the world-sheet counterpart of universality can perhaps still be studied using a manifestly supersymmetric formalism such as pure spinor formalism \cite{berkovits}.

The rest of the paper is organised as follows: The general discussion of the universality property of an arbitrary space-time theory has been given in sec.\ref{s:theorem}. Its world-sheet counterpart in the context of string theory is discussed in sec.\ref{s:ws}. Our discussion of pp-waves in CSFT has been given in sec.\ref{s:ppCSFT}. Several appendices contain various technical details.

\sectiono{A Universality in PP-Waves}
\label{s:theorem}

In this section we shall elaborate on the universality property of a space-time theory and give its precise statement. We begin by discussing the generic form of the relevant backgrounds. These are pp-waves which admit a covariantly constant null Killing symmetry. The generic ansatz for an $(N+2)$-dimensional pp-wave metric is given by \cite{brinkmann, duval91, tseytlin},
\bea
ds^2 &=&
2du dv + K(u,x) du^2 + 2 A_I(u,x) du dx^I + G^{\perp}_{IJ}(u,x) dx^I dx^J~.
\label{ppmetric}
\eea
We shall use $\mu=0, 1,\cdots ,N+1$ for the $(N+2)$-dimensional vector indices with $I=1,2,\cdots,N$ being the transverse space vector indices. The light-cone coordinates are
$u=x^+={1\over \sqrt{2}}(x^{N+1}+x^0)$ and $v=x^-={1\over \sqrt{2}}(x^{N+1}-x^0)$.
$K(u,x)$, $A_I(u,x)$ and $G^{\perp}_{IJ}(u,x)$ are certain arbitrary smooth functions. As mentioned earlier, all these functions are independent of $v$ as a requirement of the null Killing symmetry generated by: $\lt({\del\over \del v}\rt)_{\mu}$. In addition to the metric the background may contain gauge field fluxes which are null and scalars, all independent of $v$. For example, if the second rank anti-symmetric tensor\footnote{We shall denote the metric, the second rank anti-symmetric tensor gauge field, its three-form field strength and the dilaton of the low energy effective field theory of string theory by $\hat g_{\mu \nu}$, $\hat b_{\mu \nu}$, $\hat  H_{\mu \nu \rho}$ and $\hat D$ respectively.} $\hat b_{\mu \nu}$ and the dilaton $\hat D$ are switched on then the non-zero components are given as follows:
\bea
\hat H_{uIJ} &=& H_{IJ}(u,x)=\del_IB_J(u,x) -\del_JB_I(u,x)~,\quad
\hat D = \Phi(u,x)~,
\label{ppfluxscalar}
\eea
where $\hat H_{\mu \nu \rho} = \del_{[\mu} \hat b_{\nu \rho]} = \del_{\mu} \hat b_{\nu \rho}+ \del_{\nu} \hat b_{\rho \mu} +\del_{\rho} \hat b_{\mu \nu}$ is the gauge invariant flux. $H_{IJ}(u,x)$, $B_I(u,x)$ and $\Phi(u,x)$ are certain arbitrary functions independent of $v$. $\hat b_{\mu \nu}$ can therefore be chosen to have the following non-zero components:
\bea
\hat b_{uI}&=& -B_I(u,x)~.
\label{Bmunu}
\eea
Notice that for arbitrary values of the functions appearing in eqs.(\ref{ppmetric}) and (\ref{ppfluxscalar}) the background is in general off-shell.

We shall now describe what we mean by a universal sector of field configurations with a simple example. Let us consider a scalar field propagating in the background pp-wave metric (\ref{ppmetric}) with flat transverse space $G^{\perp}_{IJ}(u,x)=\delta_{IJ}$. The
lagrangian density is taken to be,
\bea
{\cal L}(\phi) = \sqrt{-\hat g}~ \phi \Box \phi~,
\label{scalaraction}
\eea
where $\hat g=\det \hat g_{\mu \nu}$. It takes the value $-1$ whenever $\hat g_{\mu \nu}$ is given by (\ref{ppmetric}) (with flat transverse space) and the laplacian computed for such a metric is,
\bea
\Box &=& {1\over \sqrt{-\hat g}} \del_{\mu} \lt[ \sqrt{-\hat g} ~\hat
  g^{\mu \nu}  \del_{\nu} \rt]~, \cr
&=& 2 \del_u \del_v + (A^IA_I-K) \del^2_v -\del_I A^I \del_v -
2 A^I\del_I\del_v + \del_{\perp}^2 ~,
\label{box}
\eea
where the transverse laplacian is $\del_{\perp}^2 = \del^I \del_I$. Notice that all the terms in the lagrangian density containing the functions $K(u,x)$ and $A_I(u,x)$, which make the background non-trivially different from flat space, also contain derivatives with respect to $v$. It is therefore immediately clear that the configurations $\phi(u,x)$, independent of $v$, effectively see the background to be flat. In other words, for all backgrounds in (\ref{ppmetric}) with $G^{\perp}_{IJ}=\delta_{IJ}$, ${\cal L}(\phi(u,x))$ is same and is equal to that computed in flat background. Therefore in the present example the universal sector of field configurations is given by the set of all $\phi(u,x)$. Notice that by restricting field
configurations to this sector we are mimicking the limit: $K\to 0$, $A_I \to 0$ and therefore universality holds even for off-shell backgrounds.

We shall now proceed to generalise the above discussion to any covariant space-time field theory which includes fields of arbitrarily high integer spins and where all the space-time fields are allowed to fluctuate. Let $\psi_{\mu \nu \cdots}(v,u,x)$ be a tensor field of arbitrary rank in a given theory. Denoting its background value for an off-shell pp-wave configuration by $\Psi_{\mu \nu \cdots}(u,x)$, we define the fluctuations $\chi_{\mu \nu \cdots}(v,u,x)$:
\bea
\psi_{\mu \nu \cdots}(v,u,x) = \Psi_{\mu \nu \cdots}(u,x) + \chi_{\mu
  \nu \cdots}(v,u,x).
\label{chi}
\eea
We shall now use notations  $\psi(v,u,x)$, $\Psi(u,x)$ and $\chi(v,u,x)$ for the sets of all possible fields $\psi_{\mu \nu \cdots}(v,u,x)$, their pp-wave background profiles and the
corresponding fluctuations respectively. If ${\cal L}(\psi(v,u,x))$ denotes the covariant lagrangian density for the present theory, then the lagrangian density for fluctuations:
\bea
{\cal L}^{\Psi}(\chi(v,u,x)) = {\cal L}(\Psi(u,x) + \chi(v,u,x))~,
\label{SPsi}
\eea
will generically depend on the background configuration $\Psi(u,x)$. But according to the universality property there exists a restricted set of fluctuations $\chi_{US}(u,x)$ independent of $v$ such that,
\bea
&&\hbox{\it Any covariant lagrangian density evaluated on }
\chi_{US}(u,x) ~\hbox{\it takes the same value} \cr
&&\hbox{\it around all the pp-waves, even off-shell, with
the same transverse space and} \cr
&& \hbox{\it same profiles for the background scalars.}
\label{universality}
\eea
i.e. if $\Psi(u,x)$ and $\Psi^{\prime}(u,x)$ are two arbitrary pp-wave backgrounds with same transverse space metric $G^{\perp}_{IJ}(u,x)$ and background scalars, then,
\bea
{\cal L}^{\Psi}(\chi_{US}(u,x)) = {\cal L}^{\Psi^{\prime}}(\chi_{US}(u,x))~.
\label{SchiUS}
\eea
In particular, if $\Psi$ corresponds to a flat transverse space with all the background scalars set to zero (or a non-zero constant for the dilaton), then ${\cal L}^{\Psi}(\chi_{US}(u,x))$ will take the same value as in flat space ${\cal L}^{flat}(\chi_{US}(u,x))$.

We shall now specify what the universal sector $\chi_{US}(u,x)$ is. First ``tensorially reduce'' a generic element $\chi_{\mu \nu \cdots}(v, u,x)$ of the set $\chi(v,u,x)$ to the transverse space. It will be decomposed into a set of fields $\chi^{(M_v,M_u)}_{IJK\cdots}(v,u,x)$, where $M_v$ and $M_u$ are the numbers of covariant $v$ and $u$ indices respectively. Then $\chi_{US}(u,x)$ is given by the set of all possible fields $\chi^{(M_v,M_u)}_{IJK\cdots}(u,x)$, independent of $v$, such that $M_u\geq M_v$.
\bea
\chi_{US}(u,x) = \lt\{\chi^{(M_v,M_u)}_{IJK\cdots }(u,x) | M_u\geq M_v \rt\}~.
\label{chiUS}
\eea
Notice that the deformations of flat background that correspond to pp-wave backgrounds (\ref{ppmetric}, \ref{ppfluxscalar}) themselves are contained in the universal sector. We prove the universality theorem (\ref{universality}) or eq.(\ref{SchiUS}) for the tensorial
fields in appendix \ref{a:argument}.

Notice that while defining $\chi_{US}$ in (\ref{chiUS}) we did not commit to any specific form of the space-time action. This automatically raises the question whether the universal sector is invariant under field redefinition or not. If we have two sets of space-time fields
$\chi(v,u,x)$ and $\hat \chi(v,u,x)$ related by a field redefinition, then a generic element in $\hat \chi(v, u,x)$ can be written as,
\bea
\hat \chi_{\mu \nu \cdots}(v,u,x) = f_{\mu \nu \cdots}(\chi(v,u,x))~,
\label{redefinition}
\eea
where the right hand side represents a generic field redefinition involving fields inside $\chi(v,u,x)$ and their derivatives. The question is if we restrict the fields on the right hand side to be inside the universal sector $\chi_{US}(u,x)$ then whether the field on
the left hand side is necessarily inside $\chi_{US}(u,x)$ or not. We have argued in appendix \ref{a:redefinition} that the answer is indeed yes.

We shall end this section with some comments about the symmetries. Certainly the isometries of the transverse space are preserved by the universal sector. But the Lorentz transformation
generated by $L^{uv}$, which is not a symmetry of any of the pp-wave backgrounds, appears to be a symmetry of the universal sector (see also \cite{mukhopadhyay06}) when the transverse space metric is independent of $u$. This transformation scales $u$ and $v$ equally in the opposite way and therefore scales the functions $K(u,x)$ and $A_I(u,x)$ while preserving the pp-wave structure of the background. But since the universal sector is insensitive to such components of the metric the transformation remains a symmetry within this sector\footnote{To see this more explicitly one notices that any term in the space-time lagrangian density evaluated on the universal sector has equal number of $u$ and $v$ indices. This is simply because the effective metric for such terms is $\eta_{\mu \nu}$ on the $(u,v)$-plane. This term therefore must be invariant under equal and opposite scaling along $u$ and $v$ directions.}. The Lorentz transformation generated by $L^{vI}$ does not preserve the pp-wave structure, though it takes the universal sector to itself. The transformation generated by $L^{uI}$ preserves the pp-wave structure, but it takes the universal sector to outside.

\sectiono{Universality on World-Sheet}
\label{s:ws}

Our discussion in the previous section has been general and applicable to any covariant field theory in space-time expanded around a pp-wave background. In this section we shall study the universality property in the context of string theory from the world-sheet point of view. In order to do so we shall consider exact pp-waves that are given by the ansatz in (\ref{ppmetric}, \ref{ppfluxscalar}) with the further restriction that the transverse space is flat and the background dilaton is independent of transverse directions:
\bea
G^{\perp}_{IJ}(u,x) = \delta_{IJ}~, \quad \Phi = \Phi(u)~.
\label{GperpPhi}
\eea
For the analysis in this section the transverse space will play the role of a spectator and therefore any $G^{\perp}_{IJ}(u,x)$ which corresponds to an on-shell background can be incorporated. We restrict it to flat space merely to simplify our presentation.

Since the statement in (\ref{universality}) concerns a space-time field theory, a natural framework to study its relevance to the world-sheet theory is string field theory. Indeed a string field theory is usually constructed for an on-shell background given by the corresponding world-sheet sigma model CFT. Using this connection here we shall derive certain world-sheet properties that are required to hold in order for the universality theorem (\ref{universality}) to be valid. In particular, we shall find that {\it the correlation functions of a specific set of world-sheet operators should be given in terms of those in flat background with both the gauge flux and dilaton switched off.} We shall verify these results
explicitly using world-sheet techniques in appendices \ref{a:Cdil} and \ref{a:NCdil}.

We begin by considering the bosonic non-polynomial CSFT action $S^{pp}$ \cite{csft, zwiebach92} around an on-shell pp-wave background (\ref{ppmetric}, \ref{ppfluxscalar}, \ref{GperpPhi}),
\bea
S^{pp}(\chi) = {1\over \kappa^2} \sum_{n=2}^{\infty} {1 \over n!} \{\chi^n\}^{pp}~,
\label{Spp}
\eea
where $\kappa$ is the closed string coupling constant which is related to the constant part of the background dilaton\footnote{We follow the notations of \cite{zwiebach92} except for the following few differences: (1) we use the notation $\chi$ for the string field instead of $\Psi$, (2) we have absorbed a factor of $\kappa$ inside the string field so that the coupling constant appears only as a pre-factor in the action, (3) we are using an additional superscript $pp$ or $flat$ on various notations to explicitly indicate the background. We have also removed the subscript $0$ from the notation $\{\cdots\}_0$ in \cite{zwiebach92}.}. $\{\chi^n\}^{pp}$ is a certain sphere correlation function in $\hbox{CFT}_{pp}\oplus \hbox{CFT}_{gh}$ which involves $n$ string fields . Here $\hbox{CFT}_{pp}$ and $\hbox{CFT}_{gh}$ denote the matter conformal
field theory corresponding to the pp-wave background and the background independent $(b,c)$ ghost CFT respectively. Among the $\{\chi^n\}$'s, $\{\chi^2 \}$ is special in the following sense: It is given by a particular inner product $\la \chi, Q^{pp} \chi\ra$, as defined in \cite{zwiebach92}, where $Q^{pp}$ is the BRST charge corresponding to the pp-wave background. Therefore the matter part of the relevant correlator involves the matter stress tensor
$(T^{pp}(z), \bar T^{pp}(\bar z))$. The matter part of any other $\{\chi^n\}$ for $n>2$ involves only the matter part of the string field. According to (\ref{universality}) the string field theory action $S_{pp}(\chi_{US})$ evaluated at a string field configuration $\chi_{US}$
that is restricted to the universal sector must be same for all pp-waves with the same dilaton profile in (\ref{GperpPhi}) (having the transverse space already been chosen to be flat)\footnote{Notice that both the statement (\ref{universality}) and its proof in appendix \ref{a:argument} consider the space-time lagrangian density while here, in the discussion of CSFT, we are considering the space-time action. This could make a difference in case of the non-constant dilaton. But it will be explained in section \ref{ss:non-constD} (see footnote \ref{calUdilaton}) that this does not happen when the background dilaton depends only on $u$, as we are considering here, and therefore our analysis with the action goes through. I thank
Ashoke Sen for emphasising this point.\label{LvsS}}. Below we shall derive the relevant sigma model correlators from this condition for the cases of constant and non-constant dilaton separately.

\subsectiono{Constant Dilaton}
\label{ss:constD}

Since the dilaton is constant and is taken to be same for all the backgrounds under consideration, the overall coupling dependent factor in eq.(\ref{Spp}) is same. Therefore in order for (\ref{universality}) to be true we must have,
\bea
\{\chi^n_{US}\}^{pp} = \{\chi^n_{US}\}^{flat}~.
\label{PsiB-Psiflat}
\eea
In order to derive the relevant CFT correlators as discussed above, let us first see what it means to have the string field restricted to the universal sector in terms of CFT. The string field can be thought of as a first quantised state in $\hbox{CFT}_{pp}\oplus \hbox{CFT}_{gh}$ satisfying certain conditions (see \cite{zwiebach92} for the complete definition of the string field). This first quantised state is expanded in terms of the basis in the relevant subspace of the matter-ghost Hilbert space. Therefore every such basis state has a matter and a ghost operator associated with it. The coefficients in this expansion turn out to play the role of the space-time fields. Since the ghosts are background independent, the tensorial properties of these fields are determined by the corresponding matter operators. Therefore the universal sector of field configurations, defined in eq.(\ref{chiUS}), corresponds to restricting the matter operators to a certain sub-sector which we call universal sector of matter operators\footnote{Notice that all the ghost operators are universal anyway as the ghost sector is background independent.}. Below we shall specify this sector explicitly. Satisfaction of eq.(\ref{PsiB-Psiflat}) for $n>2$ will then imply that all possible matter correlators involving operators only from the universal sector are universal and same as that evaluated in flat background. As discussed above, for $n=2$ there is an extra source of background dependence through the stress tensor. However, we shall see that eq.(\ref{PsiB-Psiflat}) is automatically satisfied for $n=2$ once the aforementioned universality of the relevant CFT correlators holds.

The world-sheet lagrangian density for the matter part takes the following form for the relevant backgrounds:
\bea
{\cal L}^{pp} &\sim & \del U \bar \del V + \del V \bar \del U + K(U,X)
\del U \bar
\del U + A^-_I(U,X) \del U \bar \del X^I + A^+_I(U,X) \del X^I \bar
\del U \cr && + \del X_I \bar \del X^I~,
\label{Lpp}
\eea
where $A^{\pm}_I=A_I \pm B_I$ and $I=1,2,\cdots, 24$. For an exact background this is a conformal field theory with central charge $26$. The universal sector of matter operators ${\cal O}_{US}$, which corresponds to the universal sector of field configurations given in eq.(\ref{chiUS}), is given by,
\bea
{\cal O}_{US} = \lt\{
\prod_{i=1}^{M_u} \del^{m_i} U
\prod_{\bar i=1}^{M_{\bar u}}
\bar \del^{\bar m_{\bar i}} U
\prod_{j=1}^{M_v} \del^{n_j} V
\prod_{\bar j=1}^{M_{\bar v}}
\bar \del^{\bar n_{\bar j}} V
{\cal O}(X) e^{ipU}(z,\bar z) | M_u+M_{\bar u} \geq M_v+M_{\bar v} \rt\}~,
\label{OUS}
\eea
where ${\cal O}(X)$ is an arbitrary operator in the transverse part of the CFT. In flat space these operators are well defined whose correlators can be computed easily. Although it is difficult to solve the relevant CFT for an arbitrary exact pp-wave, as has already been explained, the correlation functions of the operators inside ${\cal O}_{US}$ are universal and should be same as that evaluated in flat background,
\bea
\la \prod_i {\cal O}_i(z_i,\bar z_i) \ra^{pp} = \la \prod_i {\cal
  O}_i(z_i,\bar z_i) \ra^{flat}~, \quad \hbox{ for } {\cal O}_i \in
    {\cal O}_{US}~, \forall i~,
\label{corr-const-dil}
\eea
where $\la \cdots \ra^{pp}$ and $\la \cdots \ra^{flat}$ denote matter correlators in pp-wave and flat background without any flux respectively. We have derived  eq.(\ref{corr-const-dil}) using the world-sheet theory in appendix \ref{a:Cdil}.

We shall now show that the condition (\ref{corr-const-dil}) is enough to guarantee the relation (\ref{PsiB-Psiflat}) for $n=2$. We first write,
\bea
{\cal L}^{pp} = {\cal L}^{flat} + \Delta {\cal L}^{pp}~,
\label{DLpp}
\eea
where, ${\cal L}^{flat} \sim \del U \bar \del V +\del V \bar \del U + \del X_I \bar \del X^I$  is the lagrangian density corresponding to the flat background. Therefore the stress tensor and the BRST charge also take the similar structure:
\bea
T(z) = T^{flat}(z) + \Delta T^{pp}(z)~, \quad
\bar T(\bar z) = \bar T^{flat}(\bar z) + \Delta \bar T^{pp}(\bar z)~,
\quad Q^{pp}=Q^{flat}+\Delta Q^{pp}~,
\eea
where $(T^{flat}(z), \bar T^{flat}(\bar z))$ and $Q^{flat}$ denote the stress tensor components and the BRST charge in the flat background. Notice that $T^{flat}(z)$ and $\bar T^{flat}(\bar z)$ are contained in ${\cal O}_{US}$. This means $\la \chi_{US}, Q^{flat} \chi_{US}\ra^{pp}$ is universal. Therefore in order for eq.(\ref{PsiB-Psiflat}) to be true for $n=2$ we need to have,
\bea
\la \chi_{US}, \Delta Q^{pp} \chi_{US}\ra^{pp} = 0~.
\label{DeltaQppzero}
\eea
To see why this is right first notice that the only matter operators that appear in $\Delta Q^{pp}$ are $\Delta T^{pp}(z)$ and $\Delta \bar T^{pp}(\bar z)$ as the BRST charge is linear in the energy-momentum tensor. The next step is to argue that the operators $\Delta T^{pp}(z)$ and $\Delta \bar T^{pp}(\bar z)$ are contained in ${\cal O}_{US}$. This can be easily seen by Fourier expanding the functions $K$ and $A^I$ appearing in $\Delta T^{pp}_{\alpha \beta}$ ($\alpha$, $\beta =0,1$ referring to the standard $(\tau, \sigma)$ coordinates on the cylinder.),
\bea
\Delta T^{pp}_{\alpha \beta} &=& \lt( \del_{\alpha}U\del_{\beta}U -
       {1\over 2}\delta_{\alpha \beta} \del_{\gamma}U \del^{\gamma}U
       \rt) \int dk_u d^{24}k ~\tilde K(k_u,k) e^{ik_uU+i\vec k. \vec
     X} \cr
&& + \lt(\del_{\alpha} U\del_{\beta}X^I +\del_{\alpha}X^I
       \del_{\beta}U -\delta_{\alpha \beta}
       \del_{\gamma}U\del^{\gamma}X^I\rt) \int dk_u d^{24}k ~\tilde
       A_I(k_u,k) e^{ik_uU+i\vec k. \vec X}~, \cr &&
\label{DeltaTpp}
\eea
where $\tilde K(k_u,k)$ and $\tilde A_I(k_u,k)$ are the Fourier transforms of $K(u,x)$ and $A_I(u,x)$ respectively. Finally notice that any given term in these operators has no dependence on $V$ and at least one (first order derivative of) $U$. Then using the flat space result that a correlator involving the elements of ${\cal O}_{US}$ is nonzero only when the total numbers of $U$'s and $V$'s (we mean derivatives of them) are same, we arrive at eq.(\ref{DeltaQppzero}).

\subsectiono{Non-Constant Dilaton}
\label{ss:non-constD}

The situation is more complicated when the dilaton varies in space-time. In this case using the general argument of appendix \ref{a:argument} one has the following form of the space-time action evaluated on the universal sector:
\bea
S^{pp}(\chi_{US}) = {V_v \over \kappa^2} \int du d^{24}x ~e^{-2
  \Phi(u)} {\cal U}(u,x)~,
\label{SppUS}
\eea
where $V_v$ is the infinite volume resulting from the integration over $v$. $\Phi(u)$ denotes the background dilaton in (\ref{GperpPhi}) that appears in the world-sheet sigma model such that it does not include any constant part. This constant part is absorbed into $\kappa$ so that the effective string coupling is: $\kappa_{eff}(u)= \kappa e^{\Phi(u)}$. ${\cal U}(u,x)$ is the universal contribution, implying that\footnote{Referring the reader to the issue raised in footnote \ref{LvsS}, we notice that considering the space-time action instead
  of the lagrangian density could make a difference in the present
  case only if there were a term in ${\cal U}$ which is a total
  derivative with respect to $u$. Such terms, however, are not allowed due to general covariance. In fact, according to the general argument of appendix \ref{a:argument}, ${\cal U}(u,x)$ can involve only transverse derivatives (see footnote \ref{tr-deriv}). This is also responsible for the fact that in the present case ${\cal U}(u,x)$ does not have any dependence on the background dilaton. \label{calUdilaton}},
\bea
S^{flat}(\chi_{US}) = {V_v \over \kappa^2} \int du d^{24}x  ~{\cal U}(u,x)~.
\label{SflatUS}
\eea
We shall now translate the requirement of universality as given by eqs.(\ref{SppUS}, \ref{SflatUS}) into condition on the CFT correlators. This will be done by an induction method and therefore we begin by considering an example below.

Let us consider the graviton $\tilde h_{\mu \nu}$ in the string field
expansion:
\bea
|\chi \ra \rightarrow \int d^{26}k ~\tilde h_{\mu \nu}(k) c_1 \bar c_1
(\alpha^{\mu}_{-1} \bar \alpha^{\nu}_{-1} + \alpha^{\nu}_{-1} \bar
\alpha^{\mu}_{-1}) |k\ra~,
\eea
where $|k\ra \propto e^{ik.X}(0,0) |0\ra$, $|0\ra$ being the SL(2,C) vacuum\footnote{The CFT states $|0\ra$ and $|k\ra$ have been denoted by $|\hbox{1}\ra$ and $|\hbox{1},k\ra$ respectively in \cite{zwiebach92}.}. This contributes to $\{\Psi^2\}^{flat}$ a term of the following form (see, for example, \cite{ghoshal-sen}):
\bea
\int d^{26}k d^{26}p ~p^2 \delta(k+p) \tilde h^{\mu \nu}(k) \tilde
h_{\mu \nu}(p) = -\int {d^{26}x \over (2\pi)^{26}} ~h^{\mu \nu}(x) \del^2 h_{\mu \nu}(x)~,
\eea
where $h_{\mu \nu}(x)$ and $\tilde h_{\mu \nu}(k)$ are Fourier transform of each other.
Restricting this term to the universal sector one gets,
\bea
-{V_v \over (2\pi)^{26}} \int du d^{24} x \lt[ h^{\mu \nu}(u,x) \del_{\perp}^2 h_{\mu \nu}(u,x)\rt]_{US}~,
\label{termx}
\eea
where $[\cdots]_{US}$ represents the quantity $[\cdots ]$ with the field components restricted according to (\ref{chiUS}). Fourier transforming this universal term back to the momentum space one gets,
\bea
{V_v \over 2\pi} \int dp_u d^{24}p ~\int dk_u d^{24}k \lt\{\vec k^2 \delta(p_u+k_u)
\delta^{24}(\vec p + \vec k) \rt\}
\lt[\tilde h^{\mu \nu}(p_u,\vec p) \tilde h_{\mu \nu}(k_u, \vec k)\rt]_{US}~.
\label{termk}
\eea
In an arbitrary pp-wave background with background dilaton $\Phi(u)$ the universal term in (\ref{termx}) is supposed to be replaced by,
\bea
-{V_v \over (2\pi)^{26}} \int du d^{24} x ~e^{-2\Phi(u)} \lt[ h^{\mu \nu}(u,x)
  \del_{\perp}^2 h_{\mu \nu}(u,x)\rt]_{US}~,
\label{termPhix}
\eea
Fourier transformation of this expression gives,
\bea
&& {V_v \over 2\pi} \int dp_u d^{24}p ~\int dk_u d^{24}k
\lt\{ \vec k^2 \sum_{N=0}^{\infty} {(-2)^N\over N!}
\lt(\prod_{a=1}^N dl_{(a)} \Phi(l_{(a)}) \rt)
\delta(p_u+k_u+\sum_{a=1}^N l_{(a)}) \delta(\vec p + \vec k) \rt\} \cr
&& \lt[ \tilde h^{\mu \nu}(p_u,\vec p)
\tilde h_{\mu \nu}(k_u, \vec p)\rt]_{US}~, \cr &&
\label{termPhik}
\eea
where $\tilde \Phi(l)$ is the Fourier transform of $\Phi(u)$. Comparing the factors kept in the curly brackets in the expressions (\ref{termk}) and (\ref{termPhik}) we conclude that in presence of the non-trivial dilaton the particular two point vertex changes in the following way:
\bea
\vec k^2 \delta(p_u+k_u) \delta^{24}(\vec p + \vec k) \rightarrow
\vec k^2 \sum_{N=0}^{\infty} {(-2)^N\over N!}
\lt(\prod_{a=1}^N dl_{(a)} \Phi(l_{(a)}) \rt)
\delta(p_u+k_u+\sum_{a=1}^N l_{(a)}) \delta(\vec p + \vec k) ~. \cr
\eea
Generalising this to an $(n+2)$-point vertex one gets:
\bea
&& f(p, k, k^{(i)})\delta(p_u+k_u+\sum_{i=1}^n
k_u^{(i)}) \delta^{24}(\vec p+\vec k +\sum_{i=1}^n \vec k^{(i)}) \cr
&\rightarrow &
f(p, k, k^{(i)})
\sum_{N=0}^{\infty} {(-2)^N\over N!} \int \lt( \prod_{a=1}^N
dl_{(a)} \tilde \Phi(l_{(a)}) \rt) \cr
&& \delta(p_u+k_u+\sum_{i=1}^n k_u^{(i)}+ \sum_{a=1}^N l_{(a)})
\delta^{24}(\vec p+\vec k+\sum_{i=1}^n \vec k^{(i)})~, \cr &&
\label{term-gen}
\eea
where it is understood that each of the momenta appearing in the argument of the factor
$f(p, k, k^{(i)})$ has zero covariant $v$ component. Notice also that the above equation is schematic in the sense that depending on the vertex $f$ can carry space-time indices. For the two-point vertex considered above $f$ is a scalar as the gravitons are contracted between each other. But there are terms where the gravitons are contracted with momenta. In such cases $f$ will carry the necessary indices. Since the vertex is computed from a CFT correlator we conclude from the expression (\ref{term-gen}) that a generic matter correlator that appears in the computation of $S^{pp}(\chi_{US})$ with a non-constant dilaton is given by,
\bea
&& \la I \circ \lt[{\cal O}_{out}e^{ik_uU+i\vec k.\vec X}(0,0)\rt]
\lt(\prod_{i=1}^n{\cal O}_i e^{ik^{(i)}_uU+i\vec k^{(i)}.\vec
X}(z_i,\bar z_i)\rt) {\cal O}_{in}e^{ip_uU+i\vec p.\vec X}(0,0)
\ra_{\Phi(u)}^{pp} =\cr
&& \sum_{N=0}^{\infty} {(-2)^N\over N!} \int
\lt(\prod_{a=1}^N  dl_{(a)} \tilde \Phi(l_{(a)})\rt)\cr
&& \la I\circ \lt[ {\cal O}_{out}e^{i(k_u+\sum_{a=1}^N l_{(a)})U +i
    \vec k.\vec X}(0,0)\rt]\lt(\prod_{i=1}^n{\cal
O}_i e^{ik^{(i)}_uU+i\vec k^{(i)}.\vec X}(z_i,\bar z_i)\rt)
{\cal O}_{in}e^{ip_uU+i\vec p.\vec X}(0,0) \ra_{\Phi=0}^{flat}~, \cr &&
\hbox{ for } {\cal O}_{out}~, {\cal O}_{in}~, {\cal O}_i \in {\cal
  O}_{US} ~, \forall i~.
\label{corrPhi(u)}
\eea
where $I$ is the inversion map: $I(z)=1/z$. The correlator on the left side is computed
in $\hbox{CFT}_{pp}$, where the background contains a non-constant dilaton $\Phi(u)$. The correlators on the right side are computed in $\hbox{CFT}_{flat}$ with the dilaton set to zero. Notice that the right hand side can be computed in any other relevant background with
$\Phi=0$ as the computation is done in the universal sector. To check a simple consistency condition: if the left hand side is computed in a background with constant dilaton $\Phi(u)=c$, then one sets $\tilde \Phi(k)=c\delta(k)$ on the right side. The result turns out to be
$e^{-2c}$ times the same correlator computed with $\Phi(u)=0$, which is expected. Notice also that eq.(\ref{corrPhi(u)}) is claimed only for the universal sector - generically the result will be complicated if we sprinkle around ``non-universal" operators inside the correlator. The derivation of the result (\ref{corrPhi(u)}) using path integral has been presented in appendix \ref{a:NCdil}.

Notice that the above argument is based on the universality of ${\cal U}(u,x)$ in the sense of eqs.(\ref{SppUS}, \ref{SflatUS}) and we have not explicitly addressed the potential background dependence of the CSFT quadratic term through the energy-momentum tensor as we did in the case of constant dilaton. In presence of a non-trivial dilaton $\Phi(u)$, the energy-momentum tensor receives a correction \cite{banks86} of the following form in addition to that in eq.(\ref{DeltaTpp}) over its flat-space expression:
\bea
\Delta_{\Phi} T^{pp}_{\alpha \beta} &\propto&  \del_{\alpha} \del_{\beta} \Phi(U) - \delta_{\alpha \beta} \del^2 \Phi(U)~, \cr
&=& (\delta_{\alpha \beta} \del^2 U - \del_{\alpha} \del_{\beta}U) \int dl~ l^2 \tilde \Phi(l) e^{ilU}~,
\label{DeltaPhiTpp}
\eea
This operator belongs to ${\cal O}_{US}$ and is independent of $V$. Therefore by the same argument as given below eq.(\ref{DeltaTpp}), a matter correlator involving $\Delta_{\Phi} T^{pp}_{\alpha \beta}$ and an arbitrary number of operators inside ${\cal O}_{US}$ is zero in flat space with constant dilaton. Then using eq.(\ref{corrPhi(u)}) we conclude,
\bea
\la \chi_{US}, \Delta_{\Phi} Q^{pp} \chi_{US} \ra^{pp}_{\Phi(u)} =0~,
\eea
where $\Delta_{\Phi} Q^{pp}$ is the dilaton contribution to the BRST operator $Q^{pp}$ which is linear in $\Delta_{\Phi} T^{pp}_{\alpha \beta}$. This shows that there is no dependence of the CSFT action on the background dilaton inside the universal sector that might be introduced through the BRST operator. This is consistent with the fact that ${\cal U}(u,x)$ is independent of the background dilaton, as claimed in footnote \ref{calUdilaton}.

\subsectiono{Generalisation to NSR Superstrings}
\label{ss:nsr}

The above discussion in bosonic string theory can be generalised to the common NS sector of the Neveu-Schwarz-Ramond (NSR) superstrings. Here we demonstrate this by considering type II string theory. We have already discussed the bosonic sector of the world-sheet theory. The fermionic sector of the world-sheet lagrangian density for an exact NS-NS pp-wave takes the following form in conformal gauge,
\bea
{\cal L}^{pp}_{II} = {\cal L}^{flat}_{II}+ \Delta {\cal L}^{pp}_{II}~,
\eea
where the lagrangian density for flat background is,
\bea
{\cal L}^{flat}_{II} \sim \psi^u \bar \del \psi^v + \psi^v \bar \del \psi^u + \psi_I \bar \del \psi^I + a.h. ~,
\label{LppII}
\eea
and,
\bea
\Delta {\cal L}^{pp}_{II} &\sim & \lt[ K(U,X) \psi^u \bar \del \psi^u +
A_I(U,X) (\psi^u \bar \del \psi^I + \psi^I \bar \del \psi^u) \rt. \cr
&& \lt. + C_I(U,X) \bar \del U \psi^u \psi^I + D_{IJ}(U,X) \bar \del U \psi^I \psi^J +
E_{IJ}(U,X) \bar \del X^I \psi^u \psi^J + a.h. \rt] \cr
&& + P_{IJ}(U,X) \psi^u \psi^I \bar \psi^u \bar \psi^J
+ Q_{IJK}(U,X) (\psi^u \psi^I \bar \psi^J \bar \psi^K + \psi^J \psi^K \bar \psi^u \bar \psi^I)~,
\label{DeltaLppII}
\eea
where $C_I(u,x)$, $D_{IJ}(u,x)$, $E_{IJ}(u,x)$, $P_{IJ}(u,x)$, $Q_{IJK}(u,x)$ are certain functions constructed out of $H_{IJ}(u,x)$ and derivatives of $K(u,x)$ and $A_I(u,x)$. $a.h.$ in eqs.(\ref{LppII}, \ref{DeltaLppII}) are obtained by replacing: $\psi \to \bar \psi$, $\bar \del \to \del$ and $H_{IJ} \to - H_{IJ}$. In this case we consider the space-time tensors coming from the NS-NS sector states. The space-time indices on these fields come from both world-sheet bosons and fermions in the matter sector. Following eq.(\ref{chiUS}) one finds that the universal sector of matter operators is given by,
\bea
{\cal O}_{US}^{II} &=& \lt\{ {\cal O}^B(M^B_u,M^B_{\bar u}, M^B_v,
M^B_{\bar v})  {\cal O}^F(M^F_u,M^F_{\bar u}, M^F_v,
M^F_{\bar v}) |M_+\geq M_- \rt\} ~,
\label{OUSII}
\eea
where the bosonic part ${\cal O}^B(M^B_u,M^B_{\bar u}, M^B_v, M^B_{\bar v})$ is given by the same operator as that appears inside the curly bracket in eq.(\ref{OUS}), the only differences being that the dimensionality of the transverse space is now 8, instead of 24 and a change in notation where we have added a superscript $B$ to various numbers $M_u$ etc. appearing in eq.(\ref{OUS}). The fermionic part is given by,
\bea
{\cal O}^F(M^F_u,M^F_{\bar u},M^F_v,M^F_{\bar v})= \prod_{i=1}^{M^F_u} \del^{m_i} \psi^u
\prod_{\bar i=1}^{M^F_{\bar u}} \bar \del^{\bar m_{\bar i}} \bar \psi^u \prod_{j=1}^{M^F_v} \del^{n_j} \psi^v \prod_{\bar j =1}^{M^F_{\bar v}} \bar \del^{\bar n_{\bar j}} \psi^v
{\cal O}^F(\psi_{\perp}, \bar \psi_{\perp})(z,\bar z)~,
\label{allowed-fermion}
\eea
where ${\cal O}^F(\psi_{\perp}, \bar \psi_{\perp})$ is an arbitrary operator constructed out of the transverse $\psi^I$ and $\bar \psi^I$. The numbers $M_{\pm}$ appearing in eq.(\ref{OUSII}) are defined as follows:
\bea
M_+ &=& M^B_++M^F_+~, \quad M^B_+=M^B_u+M^B_{\bar u} ~,
\quad M^F_+=M^F_u+M^F_{\bar u} ~, \cr
M_- &=& M^B_-+M^F_-~, \quad M^B_-=M^B_v+M^B_{\bar v} ~,
\quad M^F_-=M^F_v+M^F_{\bar v} ~,
\eea
The general arguments leading to the conditions on the correlation functions in the bosonic case, namely eq.(\ref{corr-const-dil}) for constant dilaton and eq.(\ref{corrPhi(u)}) for non-constant background dilaton, remain the same. Therefore the analogues of those those conditions are given by the same equations, i.e. (\ref{corr-const-dil}) and (\ref{corrPhi(u)}) with the universal sector of operators now given as in eq.(\ref{OUSII}). For constant dilaton the  world-sheet derivation of this condition has been given in appendix \ref{a:Cdil}. For non-constant dilaton it goes through in the same way as discussed in appendix \ref{a:NCdil} for the bosonic case as any non-triviality of this analysis involves only the bosonic sector of the world-sheet theory.

\sectiono{PP-Waves in Closed String Field Theory}
\label{s:ppCSFT}
It was shown in \cite{horowitz90} that there is a large class of pp-waves that are exact solutions in string theory to all orders in $\alpha^{\prime}$ corrections. For such solutions $H_{IJ}$ in eq.(\ref{ppfluxscalar}) and
\bea
F_{IJ}(u,x) = \del_I A_J(u,x) - \del_J A_I(u,x)~,
\label{FIJ}
\eea
are taken to be functions of only $u$ and therefore are more restricted than what we have considered in (\ref{ppmetric}, \ref{ppfluxscalar}, \ref{GperpPhi}). With this restricted ansatz all the $\alpha^{\prime}$ corrections to the low energy effective field theory (EFT) equations of motion are zero. As we shall briefly review below, this is not true for our more general ansatz. In fact, in our case such unknown corrections are, in principle, non-zero at arbitrary higher orders.

In this section we shall discuss a general method of identifying an off-shell pp-wave background of EFT in CSFT with all the massive modes integrated out by their equations of motion. The off-shell ansatz that we consider is given on the EFT side by eqs.(\ref{ppmetric}), ({\ref{ppfluxscalar}) and (\ref{GperpPhi}). In general such an exact identification (certainly up to field redefinition) of off-shell field configurations is difficult to do in practice. This is because the field redefinition between EFT and CSFT with all the massive modes integrated out is in general complicated. We shall argue that for the off-shell pp-wave ansatz
that we are considering here this field redefinition is at most quadratic in fields and therefore can be found with limited computations. In this work we shall not perform any explicit CSFT computation, rather we shall formulate the problem in principle and indicate how the explicit construction can be done by computing a set of cubic terms in the action (\ref{Spp}). Our analysis will indicate that all the $\alpha^{\prime}$ corrections to the equations of motion discussed above, which will be shown to be invariant under the relevant field redefinition, can be computed using CSFT. We shall also outline an iterative method of computing the massive modes using the CSFT equations of motion in terms of the functions
$K(u,x)$, $A_I(u,x)$, $B_I(u,x)$ and $\Phi(u)$ characterising the background on the EFT side. Whenever these functions correspond to an exact CFT on the world-sheet our iterative procedure gives the corresponding CSFT solution. For definiteness we shall consider the non-polynomial bosonic CSFT action (\ref{Spp}) constructed around the flat space. But it should be possible to generalise the basic method to incorporate the common NS sectors of superstrings.

Since we shall not perform any explicit CSFT computation, our construction, as indicated above, will be based on the expectation that EFT and CSFT with all the massive modes integrated out are physically equivalent. The connection between these two theories is as follows: Integrating out all the massive fields in CSFT by satisfying their equations of motion one arrives at an effective field theory involving the same massless field content as in EFT. There could be
certain additional unphysical fields which can be eliminated away algebraically. The dynamical fields of the two theories and their off-shell gauge transformations should be related by suitable field redefinitions. Once this field redefinition is completely known any off-shell field configuration in one of the two theories can be identified in the other. This field redefinition was found at the linearised level in \cite{ghoshal-sen}, which we refer the reader to for the detailed formulation of the above connection. Below we first briefly review the off-shell pp-wave ansatz and the relevant equations of motion on the EFT side and then discuss our construction in CSFT in section \ref{ss:csft}.

\subsectiono{Effective Field Theory}
\label{ss:eft}

The field content of the EFT includes the metric $\hat g_{\mu \nu} = \eta_{\mu \nu} + \hat h_{\mu \nu}$, the dilaton $\hat D$ and the second rank antisymmetric tensor gauge field
$\hat b_{\mu \nu}$ whose three-form field strength is $\hat H_{\mu \nu \rho} = \del_{[\mu} \hat b_{\nu \rho]}$. The action is given by,
\bea
S &\propto &  \int du dv d^{24} x ~{\cal L}~, \cr
{\cal L} &=& \sqrt{-\hat g}~ e^{-2 \hat D} \lt[ R(\hat g) + 4
  \del_{\mu}\hat D \del^{\mu} \hat D - {1\over 12} \hat H_{\mu \nu
    \rho} \hat H^{\mu \nu \rho}  + {\cal O}(\alpha^{\prime}) \rt]~,
\label{EFTaction}
\eea
where the proportionality constant in the action is left arbitrary as a freedom of adding a constant piece to $\hat D$. The action receives corrections at arbitrary higher order in $\alpha^{\prime}$. The non-zero components of the off-shell background that we are interested
in are given by,
\bea
\hat h_{uu} =K(u,x) ~, \quad \hat h_{uI}=A_I(u,x) ~, \quad \hat b_{uI}=-B_I(u,x)~,
\quad \hat D = \Phi(u)~.
\label{ppwave-hathbD}
\eea
The theory possesses gauge symmetries under general coordinate transformations and anti-symmetric tensor gauge transformations. The off-shell ansatz in (\ref{ppwave-hathbD}) is not gauge invariant and therefore has been written by fixing some of the gauge symmetries. Some amount of residual gauge symmetry is still present which preserve the above ansatz. This has been discussed in appendix \ref{a:fieldredef}.

All the $\alpha^{\prime}$ corrections in (\ref{EFTaction}) are not identically zero for the generic ansatz (\ref{ppwave-hathbD}). The equations of motion reduce to the following form:
\bea
\del^I F_{IJ} =0 ~, \quad \quad \del^I H_{IJ} &=& 0~, \cr
{1\over 2} \del_{\perp}^2 K - \del_u \del^I A_I -{1\over 4}
(F_{IJ}F^{IJ} + H_{IJ}H^{IJ}) +2\del^2_u \Phi + {1\over \alpha^{\prime}} {\cal
  C}_2\lt(\sqrt{\alpha^{\prime}}\del_{\perp} F,
\sqrt{\alpha^{\prime}}\del_{\perp} H \rt) &=& 0 ~, \cr &&
\label{eeom}
\eea
where $F_{IJ}$ and $H_{IJ}$ have been defined in eqs.(\ref{FIJ}) and (\ref{ppfluxscalar}) respectively. The last equation generically receives $\alpha^{\prime}$ corrections to all
orders. All such corrections, collectively given by the term ${\cal C}_2\lt(\sqrt{\alpha^{\prime}}\del_{\perp} F, \sqrt{\alpha^{\prime}}\del_{\perp} H \rt)$, are quadratic in $F_{IJ}$ and/or $H_{IJ}$, but contain arbitrary higher powers of transverse
derivatives \cite{rudd91, tseytlin}. In a special case where both $F_{IJ}$ and $H_{IJ}$ are
independent of the transverse coordinates \cite{horowitz90}:
\bea
F_{IJ} = F_{IJ}(u)~, \quad  H_{IJ}=H_{IJ}(u)~,
\label{u-dep-FH}
\eea
all such corrections drop off. Notice that the last equation in (\ref{eeom}) is invariant under adding a purely $u$ dependent function to $K(u,x)$: $K(u,x) \to K(u,x) + f(u)$. The fact that the equations (\ref{eeom}) are at most quadratic in fields will be crucial in our following discussion on CSFT.

\subsectiono{Closed String Field Theory}
\label{ss:csft}

As mentioned earlier, here we shall formulate the problem of identifying the off-shell pp-wave configuration (\ref{ppwave-hathbD}) in CSFT. We denote the set of dynamical fields of CSFT by $\chi = \lt \{ g_{\mu \nu}=\eta_{\mu \nu} + h_{\mu \nu}, b_{\mu \nu}, D, \chi_{massive} \rt \}$, where $\chi_{massive}$ is the set of (infinite number of) massive fields. After integrating out the full set $\chi_{massive}$ one arrives at a low energy effective field theory containing only $g_{\mu \nu}$, $b_{\mu \nu}$, $D$ and certain additional non-dynamical fields that can be eliminated away algebraically. See \cite{ghoshal-sen} for the details of the procedure. The field redefinition that relates these dynamical fields and that of EFT takes the following form:
\bea
h_{\mu \nu}(v,u,x) &=& \hat h_{\mu \nu}(v,u,x) + f^h_{\mu \nu}(\hat
h_{\rho \sigma}, \hat b_{\rho \sigma}, \del_{\rho} \hat D)~,
\cr
b_{\mu \nu}(v,u,x) &=& \hat b_{\mu \nu}(v,u,x) + f^b_{\mu \nu}(\hat
h_{\rho \sigma}, \hat b_{\rho \sigma}, \del_{\rho} \hat D)~,
\cr
\del_{\mu} D(v,u,x) &=& \del_{\mu} \hat D(v,u,x) + f_{\mu}^{D}(\hat h_{\rho \sigma}, \hat b_{\rho \sigma}, \del_{\rho} \hat D)~,
\label{Psi-Psihat}
\eea
where in the last equation derivative of the dilaton has been considered in order to make the equation insensitive to a constant shift which was also left unfixed while writing down the EFT action in (\ref{EFTaction}). It was shown in \cite{ghoshal-sen} that at the linearised level the field redefinition takes the form of the above equations with all the $f$-fields set to zero. Therefore all such fields are necessarily at higher orders in the hatted fields. Notice that we have also used the well-known fact that the dilaton always appears with derivatives in these fields. We have argued in appendix \ref{a:fieldredef} that for a generic off-shell pp-wave background (\ref{ppwave-hathbD}) the only non-vanishing $f$ function is the component $f^h_{uu}(\hat h_{\rho \sigma}, \hat b_{\rho \sigma}, \hat D)$ which takes the following form:
\bea
f^h_{uu}(\hat h_{uu}=K, \hat h_{uI}=A_I, \hat b_{uI} =-B_I, \del_u \hat D
=\del_u \Phi) = {\cal K}_2(A_I, B_I, \del_u \Phi)~,
\label{fhuu}
\eea
where ${\cal K}_2(A_I,B_I,\del_u\Phi)$ is an undetermined scalar on the transverse space which is second order in the fields $A_I$, $B_I$ and $\Phi$ but may contain transverse derivatives of arbitrarily high order. Notice that such transverse derivatives can not act on $\Phi$. Therefore, by the similar argument as in appendix \ref{a:fieldredef}, the dilaton dependence is restricted only to the following type of additive terms: $\del_u \Phi \del_u \Phi$, $\del_u
\Phi \del^IA_I$ and $\del_u \Phi \del^IB_I$, where arbitrarily high powers of transverse laplacian are allowed to act on $\del^I A_I$ and/or $\del^I B_I$. This implies,
\bea
h_{uu}= K + {\cal K}_2~, \quad h_{uI}=A_I~, \quad b_{uI}=-B_I~, \quad D=\Phi~,
\label{hbD-KABPhi}
\eea
where the dilaton is identified up to an additive constant. Given these equations we shall now onwards work with the variables $h_{uu}$, $A_I$, $B_I$ and $\Phi$ on the CSFT side. One would therefore expect the CSFT equations of motion to include the first two equations in (\ref{eeom}) in the same form and the last equation in the following form:
\bea
{\cal E}_1(h_{uu}, A_I, \del_u \Phi) + {\cal E}_2(A_I, B_I, \del_u \Phi) = 0~,
\label{E1+E2}
\eea
where,
\bea
{\cal E}_1(h_{uu}, A_I, \del_u \Phi) &=& {1\over 2}\del_{\perp}^2
h_{uu} - \del_u \del^I A_I + 2\del_u^2 \Phi ~, \cr
{\cal E}_2(A_I, B_I, \del_u \Phi) &=& - {1\over 2} \del_{\perp}^2
{\cal K}_2(A_I, B_I, \del_u \Phi) - {1\over 4} (F_{IJ}F^{IJ} +H_{IJ}
H^{IJ}) \cr
&& + {1\over \alpha^{\prime}} {\cal C}_2(\sqrt{\alpha^{\prime}} \del_{\perp}F, \sqrt{\alpha^{\prime}} \del_{\perp}H)~, \cr &&
\label{E1E2}
\eea
are the terms which are linear and quadratic in fields respectively.

We shall now try to understand how the first two equations in (\ref{eeom}) and eq.(\ref{E1+E2}) are realised in CSFT. The CSFT equation of motion is given by,
\bea
Q|\chi\ra +\sum_{N=2}^{\infty} {1\over N!} |\chi^N\ra =0~,
\label{SFTeom}
\eea
where $Q=Q^{flat}$. The string field $|\chi\ra$ is expanded in terms of ghost number two states with certain properties in the combined matter-ghost conformal field theory. The fields $h_{uu}$, $h_{uI}$, $b_{uI}$ and $D$, as given by eqs.(\ref{hbD-KABPhi}), appear in the coefficients of these states in such an expansion. In particular, following \cite{ghoshal-sen} we may write for the massless component:
\bea
|\chi\ra_{massless} &=& -{1\over \sqrt{2} }
\int dk_u d\vec k \lt[ \tilde h_{uu}(k_u,\vec k)
\alpha^u_{-1} \bar \alpha^u_{-1} + \lt\{\tilde A_I(k_u,\vec k) -i
\tilde B_I(k_u,\vec k) \rt\}\alpha^u_{-1} \bar \alpha^I_{-1}\rt. \cr
&& \lt. +\lt\{\tilde A_I(k_u,\vec k) +i \tilde B_I(k_u , \vec k)
\rt\} \alpha^I_{-1} \bar \alpha^u_{-1} \rt] c_1 \bar c_1 |k_u, \vec
k\ra \cr &&
+{\sqrt{\alpha^{\prime}}\over 2} \int dk_u d\vec k ~k^I \lt\{ \tilde A_I(k_u,\vec k)
-i \tilde B_I(k_u,\vec k) \rt\} c_0^+ c_1\alpha^u_{-1}|k_u, \vec
k\ra \cr &&
-{\sqrt{\alpha^{\prime}}\over 2} \int dk_u d\vec k ~k^I \lt\{ \tilde A_I(k_u,\vec k)
+i \tilde B_I(k_u,\vec k) \rt\} c_0^+ \bar c_1 \bar \alpha^u_{-1}
|k_u, \vec k\ra
\cr &&
+\int dk_u \lt[\sqrt{2} \tilde \Phi(k_u)(c_1c_{-1} -\bar c_1 \bar c_{-1}) +
\sqrt{\alpha^{\prime}} k_u \tilde \Phi(k_u) c_0^+(c_1\alpha^u_{-1}-\bar c_1\bar
\alpha^u_{-1}) \rt] |k_u\ra~, \cr &&
\label{chi-massless}
\eea
where $\tilde h_{uu}(k_u,\vec k)$, $\tilde A_I(k_u,\vec k)$, $\tilde B_I(k_u,\vec k)$ and $\tilde \Phi(k_u)$ are the Fourier transforms of $h_{uu}(u,x)$, $A_I(u,x)$, $B_I(u,x)$ and $\Phi(u)$ respectively. $c_0^+ = {1\over \sqrt{2}}(c_0+\bar c_0)$. The first two equations in (\ref{eeom}) and the linear part in (\ref{E1+E2}) should therefore be obtained from the first term in eq.(\ref{SFTeom}). Indeed these parts can be shown to be obtained from the massless quadratic part of the action (\ref{Spp}) which was computed in \cite{ghoshal-sen}. In particular, ${\cal E}_1(h_{uu}, A_I, \del_u \Phi)$ in eq.(\ref{E1+E2}) is contributed by the following term (computed in \cite{ghoshal-sen}), in the quadratic lagrangian density (up to a
constant pre-factor):
\bea
h^{\mu \nu}\lt[{1\over 2} \del^2 h_{\mu \nu} - \del_{\mu} \del^{\rho}h_{\nu \rho} +{1\over 2} \del_{\mu} \del_{\nu} h^{\rho}_{~\rho} - \del_{\mu} \del_{\nu} D \rt] \rightarrow
h_{vv}\lt[{1\over 2} \del_{\perp}^2 h_{uu} - \del_u \del^I A_I -\del_u^2 \Phi \rt]~,
\label{hvv}
\eea
where in the second step we have restricted the fields inside square brackets according to our ansatz in (\ref{ppwave-hathbD}). Notice that the term computed is outside the universal sector as it contains the component $h_{vv}$. Generically, a higher rank equation of motion,
\bea
E_*(\hbox{universal sector}) = 0~,
\label{E*}
\eea
(where the subscript $*$ indicates that the expression has covariant tensor indices) for the fields inside universal sector can not carry covariant $v$ indices. If $*$ includes covariant $u$ indices then there exists a non-universal field varying which the equation of motion (\ref{E*}) is supposed to be obtained. Therefore to get such an equation of motion one needs to compute all the terms in the action that are linear in the relevant non-universal field\footnote{This, however, is not true if the expression on the left hand side of (\ref{E*}) is a scalar. In fact, for any scalar field solution $T(x)$ that depends entirely on the transverse directions, the relevant equation of motion can be obtained without computing any non-universal term in the action. This is why such solutions are universal. In the context of the worldvolume theory on an unstable D-brane configuration in a pp-wave background with constant dilaton this argument was applied in \cite{mukhopadhyay06} to certain tachyon solutions representing lower dimensional D-branes. Gravity solutions, on the other hand, are not generically universal because of the requirement of computing non-universal terms in the action. Pp-waves themselves are examples of this. Although these solutions belong to the universal sector, the solutions themselves are not generically universal. Universality of a
given pp-wave solution $P$ implies that it can be realised as a solution in all the pp-wave backgrounds in a given universality class (a universality class is the set of all pp-waves for which the universality theorem (\ref{universality}) holds and therefore can be characterised by the transverse space and the profiles for the background scalars). This in turn implies that $P$ can be superposed with any solution representing a pp-wave belonging to this universality class. This is not true in general as the last equation in (\ref{eeom}) is not linear. Whenever the quadratic piece drops off for a given pp-wave, the solution is universal and can be superposed with any other pp-wave solution. Nevertheless it is generically true that given two arbitrary pp-waves, a third one can always be found. But the sum rule is not simple superposition, rather more complicated when the quadratic terms are non-zero.}. In the context of (\ref{hvv}) the relevant non-universal field is $h_{vv}$ and therefore to get the full form of the relevant equation of motion one needs to compute all the terms in the action (\ref{Spp}) (for flat background) that are linear in $h_{vv}$. Fortunately, from (\ref{E1+E2}, \ref{E1E2}) it is clear that such terms must be at most cubic in order. Since the quadratic terms have already been computed in \cite{ghoshal-sen}, to complete the task one needs to compute only the cubic terms in the lagrangian density which should be given by (up to a constant pre-factor):
\bea
h_{vv}(u,x){\cal E}_2(A_I(u,x), B_I(u,x), \del_u \Phi(u))~.
\label{cubic}
\eea
From this computation one should be able to obtain all the $\alpha^{\prime}$ corrections to the EFT equation of motion given by ${\cal C}_2(\sqrt{\alpha^{\prime}} \del_{\perp}F, \sqrt{\alpha^{\prime}} \del_{\perp}H)$. Also notice that the presence of an additive term
$\del_u \Phi \del_u \Phi$ in ${\cal K}_2(A_I, B_I, \del_u \Phi)$ can not be concluded from this computation. Since it is a purely $u$ dependent term, the equation of motion is insensitive to adding such a term as explained below eq.(\ref{u-dep-FH}). With this we conclude our discussion of formulating the problem of identifying the off-shell massless pp-wave configurations of EFT in CSFT.

We shall now discuss how to compute the massive fields in CSFT once the massless components are given in terms of the functions $K(u,x)$, $A_I(u,x)$, $B_I(u,x)$ and $\Phi(u)$ following eqs.(\ref{hbD-KABPhi}). From the above discussion we have concluded:
\bea
|\chi^N\ra_{massless} = 0~, \quad \forall ~N>2 ~,
\label{*-massless}
\eea
and
\bea
Q|\chi\ra_{massless} + {1\over 2} |\chi^2\ra_{massless} =0 \quad \Rightarrow
\lt\{ \begin{array}{ll}
\del^IF_{IJ}=0~, \quad \del^IH_{IJ}=0~, \cr
{\cal E}_1(h_{uu}, A_I, \Phi) + {\cal E}_2(A_I, B_I, \del_u \Phi) = 0~,
\end{array} \rt.
\label{eom-equiv}
\eea
where $|\chi^N\ra_{massless}$ denotes the massless component of $|\chi^N\ra$. For reasons to be clarified shortly, let us denote the massless component in (\ref{chi-massless}) by $|\chi_{(1)}\ra$. Then notice that the massive part of $|\chi_{(1)}^2\ra$ can still be non-zero. This switches on higher massive modes through the equation of motion (\ref{SFTeom}). Therefore the coefficients of all such higher massive states in $|\chi\ra$ must be higher order in the functions appearing in eqs.(\ref{hbD-KABPhi}). To compute such coefficients we first expand it according to the degree of order at which those functions appear,
\bea
|\chi\ra = \sum_{n=1}^{\infty} |\chi_{(n)}\ra~,
\label{order-expand}
\eea
where $|\chi_{(n)}\ra$ is the component of the string field whose expansion involves coefficients that are only $n$-th order in these functions. For example, $|\chi_{(1)}\ra = |\chi\ra_{massless}$ in eq.(\ref{chi-massless}) is linear. Subtracting the string field equation in (\ref{eom-equiv}) from (\ref{SFTeom}) we get,
\bea
\sum_{n=2}^{\infty} \lt( Q |\chi_{(n)}\ra + |\chi_{(1)}\chi_{(n)} \ra
+ {1\over 2} \sum_{m=2}^{\infty} |\chi_{(n)}\chi_{(m)}\ra \rt) +
\sum_{N=3}^{\infty}{1\over N!} |(\sum_{n=1}^{\infty}\chi_{(n)})^N\ra +
{1\over 2} |\Delta_{(2)}\ra =0~,
\label{SFTeom-massive}
\eea
where
\bea
|\Delta_{(2)}\ra = |\chi_{(1)}^2\ra - |\chi_{(1)}^2\ra_{massless}~.
\label{D2}
\eea
To solve eq.(\ref{SFTeom-massive}) iteratively we introduce a parameter $\beta$ to count the order of fields,
\bea
&& \sum_{n=2}^{\infty} \lt( \beta^n Q |\chi_{(n)}\ra + \beta^{n+1}
|\chi_{(1)}\chi_{(n)} \ra + {1\over 2} \sum_{m=2}^{\infty} \beta^{n+m}
|\chi_{(n)}\chi_{(m)}\ra \rt) \cr &&
+\sum_{N=3}^{\infty}{1\over N!} |(\sum_{n=1}^{\infty}\beta^n
\chi_{(n)})^N\ra + \beta^2 {1\over 2} |\Delta_{(2)}\ra =0~,
\label{SFTeom-massive-beta}
\eea
We shall now show that equating the coefficient of each power of $\beta$ to zero gives us a well defined iterative procedure to construct the full solution. Equating the coefficient of $\beta^2$ to zero one
gets,
\bea
Q |\chi_{(2)}\ra + {1\over 2}|\Delta_{(2)}\ra =0 ~.
\label{2nd-order}
\eea
At the third order one finds,
\bea
Q|\chi_{(3)}\ra + |\chi_{(1)}\chi_{(2)}\ra +{1\over 3!}
|\chi_{(1)}^3\ra =0~.
\label{3rd-order}
\eea
At the $N$-th order ($N\geq 4$) the equation turns out to be,
\bea
&& Q|\chi_{(N)}\ra + |\chi_{(1)}\chi_{(N-1)}\ra + {1\over 2} \sum_{n=2}^{N-2} |\chi_{(n)} \chi_{(N-n)}\ra \cr
&& +\sum_{n=3}^N \lt(\sum_{r=0}^n \sum_{n_1,n_2, \cdots ,n_r\geq 2}\rt)_{\sum_{i=1}^r n_i -r = N-n} {1\over r!(n-r)!} |\chi_{(1)}^{n-r} \chi_{(n_1)} \cdots \chi_{(n_r)}\ra = 0~,
\label{Nth-order}
\eea
where the summations in the parenthesis in the second line are constrained. The constraint is also shown in the subscript to that parenthesis. Since $|\chi_{(1)}\ra$ is known, one can use eq.(\ref{D2}) to compute $|\Delta_{(2)}\ra$ which can, in turn, be used in eq.(\ref{2nd-order}) to compute $|\chi_{(2)}\ra$. The standard way is to fix a gauge for the massive fields so that $Q$ can be inverted to solve eq.(\ref{2nd-order}). Once $|\chi_{(2)}\ra$ is known it can be used along with $|\chi_{(1)}\ra$ in eq.(\ref{3rd-order}) to solve $|\chi_{(3)}\ra$. In order to establish that this iterative method continues to hold one needs to show that to find the solution at the $N$-th order, namely $|\chi_{(N)}\ra$ using
eq.(\ref{Nth-order}) the knowledge for $|\chi_{(N+1)}\ra$ is not required. This can be easily verified by investigating eq.(\ref{Nth-order}). For example, at $N=4$ one gets,
\bea
Q|\chi_{(4)}\ra + |\chi_{(1)}\chi_{(3)}\ra + |\chi_{(2)}^2\ra +
{1\over 2} |\chi_{(1)}^2 \chi_{(2)}\ra +{1\over 4!}|\chi_{(1)}^4\ra
=0~.
\eea
Notice that this iterative method is different from a small parameter expansion. In the latter case at every stage of the iteration an equation is solved approximately by neglecting terms of higher order. In our method an equation at a given stage of the iteration is solved exactly.

\medskip
\centerline{\bf Acknowledgement}
\noindent

I am thankful to Justin R. David, Cesar Gomez, Michael B. Green, Alok Kumar and Nicholas P. Warner for useful discussion and Juan Maldacena for drawing my attention to ref.\cite{horowitz90} at a very early stage of this work. I wish to thank Sumit R. Das and Ashoke Sen for valuable discussion and their comments on preliminary drafts. Preliminary
versions of this work were presented at CERN, where a part of this work was finished, in June 2006, Purdue and Kentucky University in October 2006 and at Saha Institute of Nuclear Physics - Calcutta, Indian Institute of Science - Bangalore and Institute of Physics - Bhubaneswar in December 2006. I thank the Theoretical Physics divisions of these institutions and universities for their interest. This work was also presented in the conferences Indian String Meeting 2006, held at Puri and Field Theory Aspects of Gravity 2006, held at Goa in December 2006. I thank the organisers and participants of these conferences for their interest. This work was financially supported by PPARC.

\appendix

\sectiono{A Proof of Universality}
\label{a:argument}

Here we shall explain our argument for why the universality theorem in (\ref{universality}) holds in the pp-wave backgrounds as described in eqs. (\ref{ppmetric}) and (\ref{ppfluxscalar}). In this argument we shall be concerned only with the tensor fields in space-time for which the universal sector is given in eq.(\ref{chiUS}). In order to proceed with the argument we shall first demonstrate this with a particular form of the space-time action, namely the low energy effective field theory action for massless fields in bosonic string theory which has all the essential features relevant to our discussion. Later we shall
generalise the argument to an arbitrary covariant action.

Therefore we start with the action in (\ref{EFTaction}) for the metric, the second rank anti-symmetric tensor gauge field and dilaton. We first expand the fields around the background (\ref{ppmetric}, \ref{ppfluxscalar}),
\bea
\hat g_{\mu \nu} = G_{\mu \nu} + \hat h_{\mu \nu}~, \quad \hat b_{\mu
\nu}=B_{\mu \nu} + \hat e_{\mu \nu}~, \quad \hat D = \Phi + \hat
\phi~,
\label{hephihat}
\eea
where $G_{\mu \nu}$, $B_{\mu \nu}$ and $\Phi$ are the background fields in eqs.(\ref{ppmetric}), (\ref{Bmunu}) and (\ref{ppfluxscalar}) respectively. Therefore $\Psi(u,x)$ and $\chi(v,u,x)$, as defined below eq.(\ref{chi}), are the sets of these background fields and the hatted fields that appear as fluctuations in the above equations in the present context.  The universal sector defined in eq.(\ref{chiUS}) is therefore given by,
\bea
\chi_{US}(u,x) &=& \lt \{
\begin{array}{llll}
\hat h_{uu}(u,x)~, & \hat h_{uv}(u,x)~, &
\hat h_{uI}(u,x)~, & \hat h_{IJ}(u,x)~, \\
\hat e_{uv}(u,x)~, & \hat e_{uI}(u,x)~, &
\hat e_{IJ}(u,x)~, & \phi(u,x)
\end{array}
\rt \}~.
\label{fluctuations}
\eea
The lagrangian density in (\ref{EFTaction}), once computed by restricting the fluctuations to the universal sector, takes the following form:
\bea
{\cal L} = e^{-\Phi(u,x)} {\cal U}(u,x)~,
\label{lag}
\eea
where ${\cal U}(u,x)$ is a sum of terms of the following generic form:
\bea
{\cal U}(u,x) &\to& \sqrt{-G(u,x)} ~G^{\mu_1\nu_1}
G^{\mu_2\nu_2} \cdots
G^{\mu_n\nu_n} [\cdots]_{\mu_1\nu_1\mu_2\nu_2\cdots \mu_n\nu_n}~, \cr
&=& \sqrt{G^{\perp}(u,x)} ~G^{\mu_1\nu_1} G^{\mu_2\nu_2} \cdots
G^{\mu_n\nu_n} [\cdots]_{\mu_1\nu_1\mu_2\nu_2\cdots \mu_n\nu_n}~,
\label{calU}
\eea
where $G(u,x) = \det G_{\mu \nu}$ and $G^{\perp}(u,x)=\det G^{\perp}_{IJ}$. In general the quantity inside the square bracket contains the following three types of objects:
\begin{enumerate}
\item
derivatives - some of which may act on the background inverse metric
pre-factors,
\item
background fields in (\ref{ppmetric}, \ref{ppfluxscalar}, \ref{Bmunu}) and
\item
fluctuations in (\ref{fluctuations}).
\end{enumerate}
To establish (\ref{universality}) or eq.(\ref{SchiUS}) we shall now argue that ${\cal U}(u,x)$ is universal among all the pp-wave backgrounds with the same transverse space metric $G^{\perp}_{IJ}(u,x)$ and background dilaton $\Phi(u,x)$. It is clear from equations (\ref{calU}) that this will be true if the following quantity is universal,
\bea
U(u,x) = G^{\mu_1\nu_1} G^{\mu_2\nu_2} \cdots
G^{\mu_n\nu_n} [\cdots]_{\mu_1\nu_1\mu_2\nu_2\cdots \mu_n\nu_n}~.
\label{U}
\eea
In order to establish universality of $U(u,x)$ we first tensorially reduce $[\cdots]_{\mu\nu \cdots}$ to the transverse space to get a set of tensors in the transverse space,
\bea
[\cdots]_{\mu\nu \cdots} \to \lt\{ [\cdots ]^{M_v,M_u}_{IJK\cdots} \rt\}~,
\label{[]decompose}
\eea
where $M_v$ and $M_u$ are the number of covariant $v$ and $u$ indices in $[\cdots]_{\mu\nu \cdots}$ respectively. Since none of the fields appearing in the expression of $[\cdots ]^{M_v,M_u}_{IJK\cdots}$ depends on $v$, all the $M_v$ number of $v$ indices are coming purely from fields and not from derivatives. Moreover since all the fields (both the backgrounds and fluctuations) are in the universal sector (\ref{chiUS}) we must have: $M_u \geq M_v~$ on the right hand side of (\ref{[]decompose}). But $[\cdots]^{M_v,M_u}_{IJK\cdots}$ with $M_u>M_v$ can not appear in $U(u,x)$ in eq.(\ref{U}) simply because the only non-zero components of the background inverse metric are:
\bea
G^{\mu \nu} \to G^{uv}~, G^{vv}~, G^{vI}~, G^{\perp IJ}~.
\label{Ginv}
\eea
Therefore we must have:
\bea
M_v=M_u ~,
\label{equality}
\eea
in all terms on the right hand side of eq.(\ref{U}). This is precisely the kind of contraction that erases the pp-wave structure of the metric background so that effectively it becomes,
\bea
G^{eff}_{\mu \nu} dx^{\mu} dx^{\nu} =  2 du dv +
G^{\perp}_{IJ}(u,x) dx^I dx^J~,
\eea
which is simply the flat Minkowski space when $G^{\perp}_{IJ}(u,x) = \delta_{IJ}$.

The above argument shows that the functions $K(u,x)$ and $A_I(u,x)$ in the background metric (\ref{ppmetric}) do not appear in $U(u,x)$. We shall now show that the functions $B_I(u,x)$ in the background flux (\ref{ppfluxscalar}) also do not appear in $U(u,x)$. In order to do so
we first notice that the facts that $[\cdots]_{\mu \nu \cdots}$ does not contain any derivative with respect to $v$, contains fields only in the universal sector together with eq.(\ref{equality}) imply that every individual field appearing in $[\cdots]_{\mu \nu \cdots}$ contains equal number of covariant $u$ and $v$ indices. For any gauge invariant background flux which is null, such as the one in (\ref{ppfluxscalar}), such components are zero. Therefore
the functions $B_I(u,x)$ do not appear in $U(u,x)$.

Notice that the transverse derivatives\footnote{Equation
  (\ref{equality}) and the restriction of field configurations to the
  universal sector imply that all the $M_u$ number of covariant $u$
  indices are also purely field indices so that all the derivatives
  that appear in $U(u,x)$ are along the transverse directions.\label{tr-deriv}} of the
background dilaton $\Phi(u,x)$ do appear in $[\cdots]_{\mu \nu \cdots}$. Because of this and the exponential pre-factor in eq.(\ref{lag}) we need to have the same dilaton profile for all
those pp-wave backgrounds for which the universality of the space-time action is supposed to hold.

We can now generalise the above argument to incorporate an arbitrary covariant action in space-time. In this case any generic term in the lagrangian density, once computed on the universal sector after expanding around a generic pp-wave background, will take the following form:
\bea
{\cal L}_{arbit} \to V\lt(\lt\{\Phi(u,x)\rt\}\rt) {\cal U}(u,x)~,
\eea
where $\lt\{ \Phi(u,x)\rt\}$ is the set of all the scalar fields in the theory at their background values and $V\lt(\lt\{\phi \rt\}\rt)$ is a multi-scalar potential depending on the term chosen and the theory. ${\cal U}(u,x)$ takes the same form as in eq.(\ref{calU}) except that now it contains tensor fields with arbitrarily high spins, which doe not have any effect on our argument above. Therefore the conclusion will be that the above term is computed to be same in all the backgrounds with the same transverse space and the same profiles $\lt\{ \Phi(u,x)\rt\}$ for the background scalars. This establishes the universality theorem in
(\ref{universality}) for the tensor fields.

\sectiono{Closure of Universal Sector under Field Redefinition}
\label{a:redefinition}

Here we shall argue that an arbitrary field redefinition takes the universal sector to itself. Referring the reader to the discussion around eq.(\ref{redefinition}) we would, therefore, like to show that,
\bea
\hat \chi_{\mu \nu \cdots}(u,x) \equiv f_{\mu \nu \cdots}(\chi_{US}(u,x))
\in \chi_{US}(u,x)~,
\label{field-redef}
\eea
where $\chi_{US}(u,x)$ is defined in (\ref{chiUS}). After tensorially reducing to the transverse space a generic term in $f_{\mu \nu \cdots}(\chi_{US}(u,x))$ takes the following form:
\bea
f_{\mu \nu \cdots}(\chi_{US}(u,x)) \to
f^{(N_u,N_v; M_u,M_v)}_{IJK\cdots} (u,x)~,
\eea
where $N_u$ ($N_v$) is the number of covariant $u$ ($v$) indices that are free while $M_u$ ($M_v$) is the same that are contracted. Notice that all the $(N_v+M_v)$ number of covariant $v$ indices, irrespective of whether they are contracted or not, should be purely field indices. Since all the fields that appear are in the universal sector, the total number of $u$ indices coming from such fields and possibly additional derivatives must satisfy,
\bea
N_u+M_u \geq N_v+M_v~.
\label{N+M}
\eea
On the other hand the total number of contracted covariant $u$ and $v$ indices, namely $(M_u+M_v)$, should be same as the total number of contravariant $u$ and $v$ indices coming from the factors of the background inverse-metric. Since the only non-zero components of background inverse-metric are as given in (\ref{Ginv}) we must have,
\bea
M_v\geq M_u~.
\label{Mvu}
\eea
Equations (\ref{N+M}) and (\ref{Mvu}) imply,
\bea
N_u \geq N_v~,
\eea
which in tern implies (\ref{field-redef}).

\sectiono{Correlators in Constant Dilaton}
\label{a:Cdil}

Here we shall argue that the matter correlation functions involving only the operators defined in eq.(\ref{OUS}) are universal in all the exact pp-wave backgrounds with constant dilaton. In particular, we shall show that eq.(\ref{corr-const-dil}) is correct. The correlator on the left hand side of eq.(\ref{corr-const-dil}) can be obtained from a correlator of the following form:
\bea
&& \la \prod_{i=1}^{M_u} \del^{m_i}U(z_i) \prod_{\bar i=1}^{{M_{\bar
      u}}} \bar \del^{\bar m_{\bar i}} U(\bar z_{\bar i})
\prod_{j=1}^{M_v} \del^{n_j}V(z'_j) \prod_{\bar j=1}^{{M_{\bar v}}}
\bar \del^{\bar n_{\bar j}} V(\bar z'_{\bar j}) \prod_{k=1}^N
e^{ip_kU}(w_k,\bar w_k) \hat {\cal O}(X)\ra^{pp} ~, \cr
&& \hbox{such that } M_u+M_{\bar u} \geq M_v + M_{\bar v}~,
\label{corr-rel}
\eea
where $\hat {\cal O}(X)$ is an arbitrary operator in the flat transverse part of the CFT. Below we shall argue that the correlator in (\ref{corr-rel}) is computed to be same as in flat background. In order to do so we first separate the world-sheet lagrangian density in (\ref{Lpp}) into ${\cal L}^{flat}$ and $\Delta {\cal L}^{pp}$ as indicated in eq.(\ref{DLpp}). Then we treat ${\cal L}^{flat}$ exactly and $\Delta {\cal L}^{pp}$ as interaction. This gives the following relation between the correlators in pp-wave and flat background:
\bea
\la \cdots \ra^{pp} = \la \cdots \ra^{flat} + \la \cdots \ra_c^{flat}~,
\eea
where the second term on the right hand side is a correction term given by,
\bea
\la \cdots \ra_c^{flat} = \sum_{N=1}^{\infty} {(-1)^N\over N!} \la
\cdots (\Delta S^{pp}[U,X])^N \ra^{flat}~,
\label{corr-rel-correc}
\eea
$\Delta S^{pp}[U,X]$ being the part of the world-sheet action in pp-wave corresponding to the lagrangian density $\Delta {\cal L}^{pp}$ in eq.(\ref{DLpp}). This contains integrated operators involving $K(U,X)$ and $A^{\pm}_I(U,X)$. In order to compute $\la \cdots \ra_c^{flat}$ we Fourier expand such functions and finally compute the correlators on the right hand side of (\ref{corr-rel-correc}) in flat background. For example, the first term of the infinite expansion is proportional to,
\bea
&& \int dk_u d\vec k \int d^2z \lt[ \tilde K(k_u,\vec k) \la \cdots
  \del U \bar \del U e^{ik_uU+i\vec k. \vec x}(z,\bar z)\ra^{flat}
  \rt. \cr
&& \lt. + \tilde A^+_I(k_u,\vec k) \la \cdots \del U \bar \del X^I
  e^{ik_uU+i\vec k. \vec x}(z,\bar z)\ra^{flat} + \tilde
  A^-_I(k_u,\vec k) \la \cdots \del X^I \bar \del U e^{ik_uU+i\vec
    k. \vec x}(z,\bar z)\ra^{flat}\rt]~, \cr &&
\label{N=1}
\eea
where $\tilde K(k_u,\vec k)$ and $\tilde A^{\pm}_I(k_u,\vec k)$ are the Fourier transforms of $K(U,X)$ and $A^{\pm}_I(U,X)$ respectively. Since all the correlators are computed in flat
background an $U$ can only be contracted with a $V$. The fact that none of the operators in (\ref{corr-rel}) and in $\Delta S^{pp}[U,X]$ involve momenta along $V$ implies, in turn, that all the $U$-derivatives\footnote{By $U$($V$)-derivative we mean an operator of the form $\del^m U(z)$ ($\del^m V(z)$) or $\bar \del^m U(\bar z)$ ($\bar \del^mV(\bar z)$). } have to be contracted with $V$-derivatives in order to produce a non-zero result. Since the total number of $U$-derivatives appearing in (\ref{corr-rel}) is always greater than or equal to that of $V$-derivatives and the higher $N$-rems in (\ref{corr-rel-correc}) introduce increasing number of $U$-derivatives and no $V$-derivatives, all the correction terms given in (\ref{corr-rel-correc}) are zero for the correlators in (\ref{corr-rel}). Therefore such correlators are computed to be same as in flat background. An equivalent path integral argument, which is more cumbersome, can also be constructed leading to the same conclusion. This establishes eq.(\ref{corr-const-dil}).

We shall now generalise the above argument for the NS-NS sector of type II superstrings as considered in section \ref{ss:nsr}. In particular, we shall argue that the same condition as in eq.(\ref{corr-const-dil}) holds for the operators given in eq.(\ref{OUSII}). In this case the relevant correlators can be obtained from a correlator of the following form:
\bea
&& \la \lt[DU\rt]^{M^B_+} \lt[DV\rt]^{M^B_-} \lt[D\psi^u\rt]^{M^F_+}
\lt[D\psi^v\rt]^{M^F_-} \hat {\cal O}^B(X) \hat {\cal O}^F(\psi_{\perp}, \bar
\psi_{\perp}) \prod_k e^{ip_kU} \ra^{pp}~, \cr
&& \hbox{such that } M^B_++M^F_+ \geq M^B_-+M^F_-~,
\label{corr-relII}
\eea
where we use schematic notations: $\lt[DU\rt]^{M^B_+}$ represents a product of $M^B_+$ number of operators of the form $\del^n U(z)$ or $\bar \del^m U(\bar w)$, $\lt[D\psi^u\rt]^{M^F_+}$ represents a product of $M^F_+$ number of operators of the form $\del^n \psi^u(z)$ or $\bar \del^m \bar \psi^u(\bar w)$ and similarly for $\lt[DV\rt]^{M^B_-}$ and $\lt[D\psi^v\rt]^{M^F_-}$ respectively. $\hat{\cal O}^F(\psi_{\perp}, \bar \psi_{\perp})$ is a fermionic analogue of $\hat{\cal O}^F(X)$ appearing in eq.(\ref{corr-rel}), namely an arbitrary operator constructed out of the transverse $\psi^I$ and $\bar \psi^I$. Just like in (\ref{corr-rel}), all the operators in (\ref{corr-relII}) are understood to be inserted at different points on the complex plane. In order to see why the correlator in \ref{corr-relII} should be same as that computed in flat background when the condition on the numbers $M^B_{\pm}$ and $M^F_{\pm}$, as shown in eq.(\ref{corr-relII}), is satisfied, we first notice that the interaction terms in the world-sheet lagrangian density now includes the fermionic sector given in eq.(\ref{DeltaLppII}) in addition to the bosonic one in eq.(\ref{DLpp}). The features of the fermionic interaction terms in (\ref{DeltaLppII}) that are relevant to our argument are as follows: (1) $\psi^v$ and $\bar \psi^v$ do not appear, (2) all the terms except $D_{IJ}(U,X) \bar \del U \psi^I \psi^J$ and its counterpart in $a.h.$ involve $\psi^u$ and/or $\bar \psi^u$ and (3) all the bosonic operators that appear in (\ref{DeltaLppII}) are from the universal sector. Since we treat the flat part in (\ref{LppII}) exactly, a $\psi^v$ ($\bar \psi^v$) appearing in the correlator (\ref{corr-relII}) can only be contracted with a $\psi^u$ ($\bar \psi^u$) either appearing in (\ref{corr-relII}) as an external operator or coming from the interaction terms in (\ref{DeltaLppII}). Therefore when $M^F_+=M^F_-$, none of the fermionic interaction terms contribute with the possible exceptions being $D_{IJ}(U,X) \bar \del U \psi^I \psi^J$ and its counterpart in $a.h.$. But even these terms do not contribute as can be argued in the following way: the bosonic sector in (\ref{corr-relII}) satisfies the
condition $M^B_+\geq M^B_-$. Therefore whenever a term like $D_{IJ}(U,X) \bar \del U \psi^I \psi^J$ appears in the correlation function it always increases the number of $U$-derivatives than that of $V$-derivatives. The result is zero due to the fact that in flat space this numbers should match in order to get a non-zero result.

Let us now discuss the cases when $M^F_+\neq M^F_-$. We begin with $M^F_+>M^F_-$. In this case the result is zero, matching with the flat space result, simply because there are not enough number of $\psi^v$ ($\bar \psi^v$) operators to contract with all the $\psi^u$ ($\bar
\psi^u$) operators. When $M^F_+<M^F_-$ the fermionic interaction terms could become important as the extra number of $\psi^v$ ($\psi^u$) operators can now be contracted with these terms. But since in this case we must have $M^B_+>M^B_-$, and all the bosonic operators in the
fermionic interaction terms are in the universal sector, the result is zero, again matching with the flat space result, as there are not enough number of $V$-derivatives to contract all the existing $U$-derivatives. This establishes that the correlators in (\ref{corr-relII}) are same as that evaluated in flat background.

\sectiono{Correlators in Non-Constant Dilaton}
\label{a:NCdil}

The analysis in the previous section does not go through when the CFT background contains non-constant dilaton as in this case one needs to be more careful about how to treat the Fradkin-Tseytlin term \cite{fradkin85} that couples the dilaton through the world-sheet curvature. In this appendix we shall discuss this issue and establish eq.(\ref{corrPhi(u)}). We start with the pre-gauge-fixed path integral,
\bea
\tilde W = \int {{\cal D}X^{\mu}{\cal D}g \over V_{diff \times Weyl}}
e^{-S}~,
\eea
where the sigma model action is,
\bea
S &=& {1\over 4\pi \alpha^{\prime}}\int d^2\sigma \sqrt{g}
\lt(g^{\alpha \beta}G_{\mu \nu}(X) + {i\epsilon^{\alpha \beta}\over
  \sqrt{g}} B_{\mu
\nu}(X) \rt)\del_{\alpha} X^{\mu}\del_{\beta} X^{\nu}  + S^{FT}~, \cr
S^{FT} &=& {1\over 4 \pi} \int d^2 \sigma \sqrt{g} ~{\cal R} \Phi(X) ~,
\eea
where $\epsilon^{01} =1$ and ${\cal R}$ appearing in the Fradkin-Tseytlin term $S^{FT}$ is the world-sheet curvature. Choosing the conformal gauge:
\bea
g_{ab}(\sigma) = e^{2\rho(\sigma)}\delta_{ab}~,
\eea
and introducing the diffeomorphism ghosts one arrives at $\tilde W = W W_{gh}$, where $W_{gh}= \int {\cal D}b {\cal D}c e^{-S_{gh}}$ is the $(b,c)$ ghost path integral and $W=\int {\cal D}X^{\mu} e^{-S}$ is the matter part of our interest. The matter action in conformal gauge
turns out to be,
\bea
S= S^{pp} + S^{FT}~,
\eea
where $S^{pp}$ corresponds to the lagrangian density in (\ref{Lpp}). Once the condition of Weyl invariance is satisfied by choosing the background to be on-shell the conformal factor $\rho(\sigma)$ can, in principle, be eliminated from the path integral relevant to the computation of any correlation function. However the effect of the Fradkin-Tseytlin term\footnote{In the second line of (\ref{SFT}) we have moved to the complex coordinate system: $z=e^{\tau + i\sigma}$.},
\bea
S_{FT} &=& -{1\over 2\pi} \int d^2 \sigma \del_a\del_a \rho \Phi(X)~, \cr
&=& -{1\over \pi} \int d^2 z \del \bar \del \rho \Phi(X)~,
\label{SFT}
\eea
has to be computed with an appropriate choice of $\rho(\sigma)$. Nevertheless the final result should be independent of this choice. Here our goal is to study the effect of this term on a correlation function involving operators inside the universal sector in an exact pp-wave background with nontrivial dilaton $\Phi(u)$. Below we first recall what happens for the constant dilaton and then we generalise the method to incorporate non-constant dilaton profile of our interest.

As it is well known, for constant dilaton: $\Phi=c$, (\ref{SFT}) can be computed explicitly by choosing a particular coordinate system on the sphere. For example:
\bea
ds^2 &=& \lt({A\over 1+z\bar z}\rt)^2 dz d\bar z =  \lt({A\over
1+w\bar w}\rt)^2 dw d\bar w~,\cr
z&=& {1\over w}~,
\eea
where $A$ is a constant determining the radius of the sphere and $(z,\bar z)$ is the local coordinate system around the {\it in}-state at $\tau \to -\infty$ whereas $(w,\bar w)$ ($w=1/z$) patches around
the {\it out}-state at $\tau \to \infty$. These coordinate systems
correspond to choosing:
\bea
\rho = \ln A - \ln(1+z\bar z) = \ln A - \ln\lt(w\bar w\over 1+w\bar w \rt)~.
\label{rho}
\eea
Converting the integration in eq.(\ref{SFT}) into a line integral in
$(w, \bar w)$-system one gets,
\bea
S_{FT}= {ic\over \pi}\oint_{C_{\infty}} dw {1\over w(1+w\bar w)}~,
\label{SFT-res}
\eea
where $C_{\infty}$ is a circle around the point $w=0$. This gives: $S_{FT}=-2c$, which is the correct result. As expected, the result does not depend on the choice of $\rho$. This can be checked in the following way: First add a variation $\delta \rho(z,\bar z)$ on the right hand side of eq.(\ref{rho}). Finiteness of the metric at $z\to \infty$ requires,
\bea
\lim_{z\to \infty} \delta \rho \to \hbox{ constant}~.
\eea
Therefore it does not contribute to the residue in eq.(\ref{SFT-res})
leaving the result unchanged.

Let us now consider the case of non-constant dilaton $\Phi(u)$ in an exact pp-wave background. In this case partially integrating the right hand side of (\ref{SFT}) and Fourier transforming $\Phi(U)$ to $\tilde \Phi(k)$ one can write:
\bea
S^{FT} &=& S^{FT}_{(1)} +  S^{FT}_{(2)} + S^{FT}_{(3)}~,
\eea
where
\bea
S^{FT}_{(1)}&=& - \int dk~ k \tilde \Phi(k) {1\over \pi} \oint_{\del R}
d\bar z \rho (z,\bar z) \bar \del U e^{ikU}(z,\bar z)~, \cr
S^{FT}_{(2)} &=& \int dk~ k^2 \tilde \Phi(k) {1\over \pi} \int d^2z~
\rho(z,\bar z) \del U\bar \del U e^{ikU}(z,\bar z) \cr
&& -i \int dk~ k \tilde \Phi(k) {1\over \pi} \int d^2z~ \rho(z,\bar z)
\del \bar \del U e^{ikU}(z,\bar z)~,\cr
S^{FT}_{(3)} &=& {i\over \pi}\oint_{\del R} dz~ \del \rho \Phi(U) =
\int dk~ \tilde \Phi(k) {i\over \pi} \oint_{\del R} dw~ {1\over
w(1+w\bar w)} I\circ e^{ikU}(w,\bar w)~, \cr &&
\label{SFT123}
\eea
Where $I\circ \phi$ is the conformal transformation of $\phi$ under
inversion $I(z)=1/z$. $\del R$ is the boundary of the coordinate patch
$(z,\bar z)$. We are interested in computing the following correlation function:
\bea
\la \cdots \ra^{pp}_{\Phi(u)} &=& \int {\cal D} X^{\mu} e^{-S^{pp}-S^{FT}}
\cdots~, \cr
&=& \sum_{N=0}^{\infty} {(-1)^N\over N!} \int {\cal D}X^{\mu} e^{-S^{pp}}
\lt(S^{FT}_{(1)} + S^{FT}_{(2)} + S^{FT}_{(3)}\rt)^N \cdots ~,
\label{arbit-corr}
\eea
where $\cdots$ includes operators inside the universal sector. Notice that all the operators appearing in (\ref{SFT123}) are themselves inside the universal sector. Therefore using the result of the previous appendix we conclude:
\bea
\la \cdots \ra^{pp}_{\Phi(u)} &=& \sum_{N=0}^{\infty} {(-1)^N\over N!} \la \cdots \lt(S^{FT}_{(1)} + S^{FT}_{(2)} + S^{FT}_{(3)}\rt)^N \ra^{flat}_{\Phi=0}~.
\eea
The operators in $S^{FT}_{(1)}$ and $S^{FT}_{(2)}$ necessarily involve $U$-derivatives. Since such operators can contract only with $V$-derivatives, which are not available in sufficient number, the contribution of such terms drop off from the above correlator. This implies:
\bea
\la \cdots \ra^{pp}_{\Phi(u)} &=& \sum_{N=0}^{\infty} {(-1)^N\over N!} \la \cdots \lt(S^{FT}_{(3)}\rt)^N \ra^{flat}_{\Phi=0}~.
\eea
To compute this we first notice that $S^{FT}_{(3)}$ involves a line integral along $\del R$, which is a collection of circles $C_i$'s around the insertions included in $\cdots$ which are at finite values of $(z,\bar z)$ and a circle $C_{\infty}$ around the point $z=\infty ~(w=0)$ which can also support an insertion from $\cdots$. Since $e^{ikU}$ can not be contracted with any operators in $\cdots$, contributions will come only from $C_{\infty}$. It is then straightforward to establish the result (\ref{corrPhi(u)}) by evaluating the necessary residues.

The fact that the final result does not depend on the choice of $\rho$ follows from the same reasoning as in the case of constant dilaton. This, however, is not as straightforward as presented here in case of a correlation functions with non-universal operators. In that case the contributions from $S^{FT}_{(1)}$ and $S^{FT}_{(2)}$ are not necessarily zero. Out of these two, $S^{FT}_{(2)}$ is sensitive to the local variation of the Weyl factor. The second term in its expression in (\ref{SFT123}) is zero by the equation of motion: $\del \bar \del U=0$. This holds in the full pp-wave CFT because of which one can always fix the light-cone gauge for such backgrounds. Moreover $\del \bar \del U$ does not contract with $e^{ikU}$. This is true in flat background and does not get spoiled in presence of the interactions given by $\Delta {\cal L}^{pp}$ in eq.(\ref{DLpp}). From the structure of the world-sheet operator appearing in the first term in $S^{FT}_{(2)}$ we infer that it contributes to the $uu$ component of the metric beta function. In fact it corresponds to the $\del^2_u \Phi$ term in the equation of motion in (\ref{eeom}). In the above analysis we have considered the Weyl dependence of only $S^{FT}$ and therefore this is the only contribution expected. Considering the complete Weyl dependence one expects this contribution to go away due to the space-time equation of motion. $S^{FT}_{(1)}$, on the other hand, receives contributions only from the boundaries of the punctured sphere. Such effects need to be handled with more care in a generic situation by considering the complete Weyl dependence including that of the insertions at the punctures. However, such issues can be neglected for correlation functions involving only the universal sector operators.

\sectiono{Field Redefinition and Gauge Transformation}
\label{a:fieldredef}

Here we shall first find the higher order terms in the field redefinition (\ref{Psi-Psihat}) given by the second rank tensor and vector fields: $f^h_{\mu \nu}(\hat h_{\rho \sigma}, \hat b_{\rho \sigma}, \del_{\rho} \hat D)$, $f^b_{\mu \nu}(\hat h_{\rho \sigma}, \hat b_{\rho \sigma}, \del_{\rho} \hat D)$ and $f_{\mu}^{D} (\hat h_{\rho \sigma}, \hat b_{\rho \sigma}, \del_{\rho} \hat D)$ when the off-shell field configurations are restricted to the pp-wave ansatz in eq.(\ref{ppwave-hathbD}). In particular, by using arguments similar to that in \cite{horowitz90} we shall show that the only non-zero component is as given in eq.(\ref{fhuu}). Later we shall also discuss the gauge symmetries of EFT and CSFT and how they are related to this field redefinition.

In order to find the $f$-fields appearing in eqs.(\ref{Psi-Psihat}) we first notice that all such fields should necessarily be expressible in terms of the non-zero components in the
background pp-wave ansatz in (\ref{ppwave-hathbD}), Christoffel connection and derivatives. The only non-zero components of the Christoffel connection are \cite{tseytlin},
\bea
\hat \Gamma^{\mu}_{\nu \rho} \to \hat \Gamma^v_{uu}~, \quad \hat \Gamma^v_{uI}~, \quad \hat \Gamma^v_{IJ}~, \quad \hat \Gamma^I_{uu}~,
\quad \hat \Gamma^I_{uJ}~.
\label{hatGamma}
\eea
It turns out that a covariant $u$ index can not be contracted \cite{horowitz90}. This is because the background inverse metric (\ref{Ginv}) allows contraction only with a covariant $v$ index and a non-zero field with a covariant $v$ index can not be constructed in this case. It is also not possible to construct a field without a covariant $u$ index. Because of this and the fact that all the $f$-fields are necessarily higher order in the functions appearing in (\ref{ppwave-hathbD}) we conclude that any non-zero $f$-field (1) should have at least one covariant $u$ index, (2) should have no covariant $v$ index, (3) can not have a non-zero covariant $u$ index if it is a vector. Therefore the only nonzero $f$-fields are: $f^h_{uu}$, $f^h_{uI}$ and $f^b_{uI}$. Let us first consider $f^h_{uu}$. Because of the same reasons mentioned above it must be quadratic in fields. Any higher order term necessarily has more than two covariant $u$ indices. Therefore the allowed expression for $f^h_{uu}$ is precisely given by ${\cal K}_2(A_I,B_I, del_u \Phi)$ as described below eq.(\ref{fhuu}). Using similar reasoning one can argue that $f^h_{uI}$ and $f^b_{uI}$ can at most be linear in fields and therefore must be zero.

We shall now discuss the gauge symmetries of the EFT and CSFT. Although the gauge transformations are understood on the EFT side, they are not fully known on the CSFT side. After integrating out all the massive modes the CSFT gauge transformations for the massless fields take complicated form in general. But in our case, since we know the form of the complete field redefinition, we shall find the relevant CSFT gauge transformations fully. Once the expression for ${\cal K}_2(A_I,B_I, \del_u \Phi)$ is computed through the CSFT computation as indicated in sec. \ref{ss:csft}, such gauge transformations will be completely known and it should be possible to verify them explicitly in CSFT.

We begin by discussing the gauge symmetries on the EFT side. This theory possesses gauge invariance under the anti-symmetric tensor gauge transformation (TGT) and general coordinate transformation (GCT). The infinitesimal form of TGT, parameterised by $\hat \zeta_{\mu}(u,v,x)$, is given by,
\bea
\delta_{\hat \zeta} \hat b_{\mu \nu} = \del_{\mu} \hat \zeta_{\nu} - \del_{\nu} \hat \zeta_{\mu} ~, \quad
\delta_{\hat \zeta} \hat g_{\mu \nu} = 0~, \quad \delta_{\hat \zeta} \hat D = 0~.
\label{delta-zeta}
\eea
Since the only non-zero gauge invariant background flux components are $\hat H_{uIJ} = \hat H_{IJ}(u,x)$ (see eqs.(\ref{ppfluxscalar})), the components $\hat b_{uv}$, $\hat b_{vI}$ and $\hat b_{IJ}$ should at most be pure gauge with the only remaining non-trivial components
being $\hat b_{uI}(u,x)$. The pure gauge components are set to zero by using the gauge freedom corresponding to the components $\hat \zeta_v$ and $\hat \zeta_I$. Therefore the non-trivial component of the residual gauge transformation, parameterised by: $\hat \zeta^{res}_{\mu}=(\hat \zeta_u(u,x), \hat \zeta_v=0, \hat \zeta_I=0)$, is given by:
\bea
\delta^{res}_{\hat \zeta} \hat b_{Iu} = \del_I \hat \zeta_u~.
\label{delta-zeta-res-hat}
\eea
The infinitesimal form of GCT, parameterised by $\hat \xi_{\mu}(u,v,x)$, is given by,
\bea
\delta_{\hat \xi} \hat g_{\mu \nu} &=& \del_{\mu}\hat \xi_{\nu} + \del_{\nu}
\hat \xi_{\mu} - 2 \hat \Gamma^{\rho}_{\mu \nu} \hat \xi_{\rho}~, \cr
\delta_{\hat \xi} \hat b_{\mu \nu} &=& \del_{\mu} \hat \xi^{\rho} \hat b_{\rho
  \nu} + \del_{\nu} \hat \xi^{\rho} \hat b_{\mu \rho} + \hat \xi^{\rho}
\del_{\rho} \hat b_{\mu \nu}~, \cr
\delta_{\hat \xi} \hat D &=& \hat \xi^{\rho} \del_{\rho} \hat D~,
\label{delta-xi}
\eea
A part of this gauge invariance is broken by the ansatz in (\ref{ppwave-hathbD}). The non-trivial components of the residual transformations are parameterised by: $\hat \xi^{res}_{\mu}= (\hat \xi_u(u,x), \hat \xi_v=0, \hat \xi_I(u))$ and can be shown to
take the following form:
\bea
\delta^{res}_{\hat \xi} \hat h_{uu}
&=& 2 \del_u \hat \xi_u + (\del^I K - 2\del_u A^I) \hat \xi_I~, \cr
\delta^{res}_{\hat \xi} \hat h_{uI}
&=& \del_I \hat \xi_u + \del_u \hat \xi_I - F_I^{~J} \hat \xi_J~,   \cr
\delta^{res}_{\hat \xi} \hat b_{uI} &=& -\del^J B_I \hat \xi_J ~,
\label{delta-xi-pp}
\eea
The residual gauge transformations in (\ref{delta-zeta-res-hat}, \ref{delta-xi-pp}) should leave the equations of motion (\ref{eeom}) invariant. It is straightforward to see this for (\ref{delta-zeta-res-hat}). The transformation in (\ref{delta-xi-pp}) implies,
\bea
&& \delta^{res}_{\hat \xi} (\del_{\perp}^2 K - 2 \del^I \del_u A_I) = \del^K
(\del_{\perp}^2 K - 2 \del^I \del_u A_I) \hat \xi_K~, \cr
&& \delta^{res}_{\hat \xi} F_{IJ} = \del^K F_{IJ} \hat \xi_K ~, \quad
\delta^{res}_{\hat \xi} H_{IJ} = \del^K H_{IJ} \hat \xi_K ~,
\label{delta-xi-pp2}
\eea
Using these transformations one can show that, as expected, the variation of the equations of
motion in (\ref{eeom}) is zero. It is straightforward to see this for the first two equations. The variation of the left hand side of the last equation in (\ref{eeom}) gives,
\bea
\hat \xi_K \del^K \lt[
{1\over 2} \del_{\perp}^2 K - \del_u \del^I A_I -{1\over 4}
(F_{IJ}F^{IJ}+ H_{IJ}H^{IJ}) + {1\over \alpha^{\prime}}{\cal
  C}_2\lt(\sqrt{\alpha^{\prime}}\del_{\perp} F,
\sqrt{\alpha^{\prime}}\del_{\perp} H \rt) \rt]~.
\eea
By the same equation of motion, the expression inside the square bracket is dependent only on $u$ and therefore the result is zero.

The residual gauge invariance in (\ref{delta-zeta-res-hat}, \ref{delta-xi-pp}) and the field redefinition (\ref{Psi-Psihat}, \ref{fhuu}) determine the form of the gauge transformations on the CSFT side. We shall now discuss this systematically and finally come up with certain results that need to be verified with explicit CSFT computations, which we leave for future work. The infinitesimal gauge transformation of the full CSFT is given by:
\bea
\delta |\chi \ra = Q |\Lambda \ra + \sum_{N=3}^{\infty} {1\over (N-2)!} |\chi^{N-2} \Lambda \ra~,
\label{csft-gt}
\eea
where $|\Lambda \ra$ is the ghost number one infinitesimal gauge transformation parameter and we have absorbed a factor of $\kappa$ in it with respect to \cite{zwiebach92}. The above equation encodes the gauge transformation of all the higher massive fields. We are, however, interested in the transformations of the dynamical massless components only once all the massive ones have been integrate out by using their equations of motion. Generically this procedure leads to a complicated form of the gauge transformation for the massless fields on the CSFT side. It was shown in \cite{ghoshal-sen} that taking,
\bea
|\Lambda \ra = -{i\over \sqrt{2}}\int d^{26}k \lt[ i\tilde \zeta_{\mu} (k) (c_1
\alpha^{\mu}_{-1} + \bar c_1 \bar \alpha^{\mu}_{-1} + \sqrt{\alpha^{\prime}\over
  2} k^{\mu} c^+_0 ) - \tilde \xi_{\mu}(k)
(c_1\alpha^{\mu}_{-1} - \bar c_1 \bar \alpha^{\mu}_{-1})\rt] |k\ra~,
\eea
and interpreting $\tilde \zeta$ and $\tilde \xi$ to be the Fourier transforms of $\hat \zeta$
and $\hat \xi$ respectively the gauge transformation in (\ref{csft-gt}) reproduces the gauge transformations in (\ref{delta-zeta}, \ref{delta-xi}) at the linearised level. It means that, as expected, the complications come from the higher order terms in ({\ref{csft-gt}). Therefore the general form of the relevant gauge transformations can be written as:
\bea
\delta_{\zeta} b_{\mu \nu} = \del_{\mu} \zeta_{\nu} - \del_{\nu}
\zeta_{\mu} + \Delta^{(b)}_{\mu \nu}~, \quad
\delta_{\zeta} g_{\mu \nu} = \Delta^{(g)}_{\mu \nu} ~, \quad
\delta_{\zeta} D = \Delta^{(D)}~, \cr
\delta_{\xi} b_{\mu \nu} = \Theta^{(b)}_{\mu \nu} ~, \quad \delta_{\xi} g_{\mu \nu} = \del_{\mu} \xi_{\nu} + \del_{\nu} \xi_{\mu} + \Theta^{(g)}_{\mu \nu} ~, \quad \delta_{\xi} D= \Theta^{(D)}~,
\label{delta-zeta-xi-csft}
\eea
where all the higher order terms given by $\Delta$'s and $\Theta$'s are linear in $\zeta_{\mu}$ and $\xi_{\mu}$ respectively, but can depend on the massless fields $h_{\mu \nu}$, $b_{\mu \nu}$ and $\del_{\mu} D$ to arbitrarily high order. The gauge transformation parameters $\zeta_{\mu}$ and $\xi_{\mu}$ are same as $\hat \zeta_{\mu}$ and $\hat \xi_{\mu}$ in (\ref{delta-zeta}, \ref{delta-xi}) respectively at the linearised level:
\bea
\zeta_{\mu} = \hat \zeta_{\mu} + \Delta^{(\hat \zeta)}_{\mu}~, \quad
\xi_{\mu} = \hat \xi_{\mu} + \Theta^{(\hat \xi)}_{\mu}~,
\label{zeta-xi}
\eea
where $\Delta^{(\hat \zeta)}_{\mu}$ and $\Theta^{(\hat \xi)}_{\mu}$ contain all the higher order terms. These contributions are also linear in the gauge transformation parameters appearing in the superscripts, but can depend on the massless fields to arbitrarily high order. However, as we shall see below, for our off-shell pp-wave ansatz the $\Delta$'s and $\Theta$'s in eqs.(\ref{delta-zeta-xi-csft}) are at most quadratic and those in eqs.(\ref{zeta-xi}) are at most linear in the massless fields.

Let us begin with TGT. The residual gauge transformation is parameterised by: $\zeta^{res}_{\mu} = \hat \zeta^{res}_{\mu} + \Delta^{(\hat \zeta^{res})}_{\mu}$, where $\hat \zeta^{res}_{\mu}$ has been given above eq.(\ref{delta-zeta-res-hat}). Since $\Delta^{(\hat \zeta^{res})}_{\mu}$ has to be linear in $\hat \zeta^{res}_{\mu}$ and at least linear in the massless fields, there is no way to construct this. Also the field redefinition relating $b_{uI}$ and $\hat b_{uI}$ is trivial in this case. Therefore the first equation in (\ref{delta-zeta-xi-csft}) reduces to,
\bea
\delta^{res}_{\zeta} b_{Iu} = \del_I \zeta_u~, \quad \zeta_u=\hat \zeta_u~.
\eea
However, for GCT the residual transformation parameter: $\xi^{res}_{\mu} = \hat \xi^{res}_{\mu} + \Theta^{(\hat \xi^{res})}_{\mu} = (\xi_u, \xi_v, \xi_I)$, where $\hat \xi^{res}_{\mu}$ is given above eq.(\ref{delta-xi-pp}), can take the following form:
\bea
\xi_u(u,x) &=& \hat \xi_u(u,x) + L_u^{~I}(u,x) \hat \xi_I(u)~, \quad \xi_v =0~, \quad \xi_I = \hat \xi_I~, 
\eea
where $L_{uI}$ is linear in $A_I$ and/or $B_I$ but may contain arbitrarily high powers of transverse laplacian. This and the field redefinition (\ref{Psi-Psihat}, \ref{fhuu}) determines the residual gauge transformation completely.
\bea
\delta^{res}_{\xi}h_{uu} &=& 2 \del_u \xi_u + (\del^IK -2\del_uA^I) \xi_I
+ {\cal K}^{(A)I} (\del_I\xi_u - \del_u \xi_I- F_I^{~J}\xi_J) + {\cal K}^{(B)I} \del^J B_I \xi_J \cr
&& -2 \del_u \lt(L_u^{~J}\xi_J\rt) - {\cal K}^{(A)I} \del_I L_u^{~J}\xi_J~, \cr
\delta^{res}_{\xi} h_{uI} &=& \del_I \xi_u + \del_u \xi_I - F_I^{~J}\xi_J - \del_I L_u^{~J}\xi_J~, \cr
\delta^{res}_{\xi} b_{uI} &=& \del^J B_I \xi_J~,
\label{delta-xi-csft}
\eea
where ${\cal K}^{(A)I}$ and ${\cal K}^{(B)I}$ are defined to be,
\bea
\delta_{A} {\cal K}_2 = {\cal K}^{(A)I} \delta A_I~, \quad
\delta_{B} {\cal K}_2 = {\cal K}^{(B)I} \delta B_I~,
\eea
and therefore are linear in massless fields. This shows that the transformations in (\ref{delta-xi-csft}) are at most quadratic in massless fields. Therefore to verify this in CSFT one needs to compute the right hand side of (\ref{csft-gt}) up to at most $N=4$. However, it was pointed out in \cite{ghoshal-sen} that a certain {\it trivial} symmetry of the CSFT action needs to be considered in addition to the pure gauge transformation in (\ref{csft-gt}) in order to realise TGT and GCT of the EFT completely at higher order. Therefore the transformations in (\ref{delta-xi-csft}) may not be obtained entirely from (\ref{csft-gt}). It would be interesting to understand this explicitly through CSFT computations.

\end{document}